\documentclass[pre,twocolumn,amsmath,amssymb,floatfix,superscriptaddress,nopacs]{revtex4}

\usepackage{graphicx}
\usepackage{latexsym}
\usepackage{amsmath}
\usepackage{amssymb}
\usepackage{amsfonts}
\usepackage{bm}
\usepackage{color}

\newcommand{\ddiff}{\ensuremath{\mathrm{d}}}

\newcommand{\rvec}{\bf r}
\newcommand{\qvec}{\bf q}
\newcommand{\la}{\left<}
\newcommand{\ra}{\right>}
\newcommand{\kB}{k_\mathrm{B}}
\newcommand{\kBT}{k_\mathrm{B}T}
\newcommand{\df}{d_f}
\newcommand{\dperi}{d_s}
\newcommand{\avM}{\la M \ra}
\newcommand{\avN}{\la N \ra}

\newcommand{\Rend}{R_\mathrm{e}}
\newcommand{\Rgyr}{R_\mathrm{g}}
\newcommand{\Ren}{R_\mathrm{e,n}}
\newcommand{\Rez}{R_\mathrm{e,z}}
\newcommand{\Rgn}{R_\mathrm{g,n}}
\newcommand{\Rgz}{R_\mathrm{g,z}}
\newcommand{\Estar}{E_{\star}}
\newcommand{\Nstar}{N_{\star}}

\newcommand{\phistarstar}{\phi_{\star\star}}
\newcommand{\DIL}{\mathrm{dil}}
\newcommand{\DEN}{\mathrm{den}}
\newcommand{\SEM}{\mathrm{sd}}
\newcommand{\alphaDIL}{\alpha_{\DIL}}
\newcommand{\alphaDEN}{\alpha_{\DEN}}
\newcommand{\alphaSEM}{\alpha_{\SEM}}
\newcommand{\deltaDIL}{\delta_{\DIL}}
\newcommand{\deltaDEN}{\delta_{\DEN}}
\newcommand{\gammaDIL}{\gamma_{\DIL}}
\newcommand{\gammaDEN}{\gamma_{\DEN}}
\newcommand{\nuDIL}{\nu_{\DIL}}
\newcommand{\nuDEN}{\nu_{\DEN}}
\newcommand{\fend}{f_{\mathrm{end}}}
\newcommand{\funknown}{f_{\mathrm{u}}}
\newcommand{\aunknown}{\alpha_{\mathrm{u}}}
\newcommand{\ficut}{f^{\mathrm{cut}}_i}

\begin{document}

\title{Conformational properties of strictly two-dimensional equilibrium polymers}

\author{J.P.~Wittmer}
\email{joachim.wittmer@ics-cnrs.unistra.fr}
\affiliation{Institut Charles Sadron, Universit\'e de Strasbourg \& CNRS, 
23 rue du Loess, 67034 Strasbourg Cedex, France}

\author{A. Cavallo}
\affiliation{Institut Charles Sadron, Universit\'e de Strasbourg \& CNRS, 
23 rue du Loess, 67034 Strasbourg Cedex, France}

\author{A. Johner}
\affiliation{Institut Charles Sadron, Universit\'e de Strasbourg \& CNRS, 
23 rue du Loess, 67034 Strasbourg Cedex, France}

\begin{abstract}
Two-dimensional monodisperse linear polymer chains are known to adopt for sufficiently 
large chain lengths $N$ and surface fractions $\phi$ compact configurations with fractal perimeters. 
We show here by means of Monte Carlo simulations of reversibly connected hard disks 
(without branching, ring formation and chain intersection) that polydisperse self-assembled 
equilibrium polymers with a finite scission energy $E$ are characterized by the same universal 
exponents as their monodisperse peers. Consistently with a Flory-Huggins mean-field approximation,
the polydispersity is characterized by a Schulz-Zimm distribution with a susceptibility exponent 
$\gamma=19/16$ for all not dilute systems and the average chain length $\avN \propto \exp(\delta E) \phi^{\alpha}$ 
thus increases with an exponent $\delta = 16/35$. Moreover, it is shown that $\alpha=3/5$ for semidilute solutions 
and $\alpha \approx 1$ for larger densities.
The intermolecular form factor $F(q)$ reveals for sufficiently large $\avN$ a generalized Porod scattering 
with $F(q) \propto 1/q^{11/4}$ for intermediate wavenumbers $q$ consistently with a fractal perimeter dimension 
$\dperi=5/4$.
\end{abstract}
\date{\today}
\maketitle


\section{Introduction}
\label{sec_intro}

%
Strictly two-dimensional ($d=2$) linear and {\em monodisperse} polymer chains are well-known 
to adopt for sufficiently large chain lengths $N$ and densities $\phi$ compact configurations 
\cite{DegennesBook,VanderzandeBook,Duplantier86a,Duplantier89,CarmesinKremer90,ANS03,CMWJB05} 
of fractal perimeter \cite{MKA09,MSZ11}.
They are characterized by the exponents $\nu$, $\gamma$, $\theta_0$, $\theta_1$ and $\theta_2$
indicated in the fourth column of Table~\ref{tab_expo}, e.g., by a Flory exponent $\nu=1/2$ 
characterizing the typical size $R \propto N^{\nu}$ of chains with respect to their mass $N$
\cite{foot_contactexp}.
With the fractal dimension $\df$ being defined by $N \simeq R^{\df}$,
$\df \equiv 1/\nu = d=2$ in this density limit. Interestingly, it is also known that 
these compact monodisperse chains do not adapt regular shapes with smooth perimeters (surfaces) 
of surface dimension $\dperi=d-1=1$.
(The surface dimension $\dperi$ of a compact object is defined by the asymptotic scaling
$S \simeq R^{\dperi}$ of its surface $S$ with respect to its size $R$ \cite{MandelbrotBook}.)
In fact, the irregular perimeters are characterized by a fractal surface exponent \cite{MKA09,MSZ11}
\begin{equation}
\dperi = \df - \theta_2 = 5/4 > 1
\label{eq_dperi}
\end{equation}
set by the known exponents $\df=2$ and $\theta_2=3/4$. 
As an experimentally measurable consequence, the intramolecular form factor $F(q)$ \cite{BenoitBook} 
is thus described by the generalized Porod scattering relation \cite{BenoitBook,MKA09,MSZ11} 
\begin{equation}
F(q)/F(0) \approx 1/Q^{2 \df -\dperi} = 1/Q^{11/4},
\label{eq_genPorod}
\end{equation}
with $Q = q R(N)$ being the reduced wavevector, rather than by the usual Porod scattering 
$F(q)/F(0) \approx 1/Q^3$ of smooth compact objects in $d=2$.
More information on the defining properties of the universal asymptotic exponents indicated 
in Table~\ref{tab_expo} will be given below.

\begin{table}[t]
\begin{tabular}{|c||c|c|c|c|c|}
\hline
           & mean  & dilute       & dense   & dilute  & dense  \\ 
           & field &  $d=2$       &  $d=2$  & $d=3$   & $d=3$  \\ \hline
$\nu$      & $1/2$ & $\bm{3/4}$   & $\bm{1/2}$   & $\approx 0.588$ & $1/2$  \\ 
$\gamma$   & $1$   & $\bm{43/32}$ & $\bm{19/16}$ & $\approx 1.165$ & $1$ \\ 
$\theta_0$ & $0$   & $\bm{11/24}$ & $\bm{3/8}$   & $\approx 0.275$ & $0$ \\ 
$\theta_1$ & $0$   & $\bm{5/6}$   & $\bm{1/2}$   & $\approx 0.45$  & $0$ \\ 
$\theta_2$ & $0$   & $\bm{19/12}$ & $\bm{3/4}$   & $\approx 0.53$  & $0$ \\ 
\hline
$\delta$   & $1/2$ & $\bm{32/75}$ & $\bm{16/35}$ & $\approx 0.462$ & $1/2$ \\ 
$\alpha$   & $1/2$ & $\bm{32/75}$ & $\bm{3/5}$   & $\approx 0.462$ & $\approx 0.67$ \\ 
\hline
\end{tabular}
\caption[]{Summary of theoretically predicted asymptotic exponents characterizing either 
dilute and dense solutions of long linear polymer in $d=2$ and $d=3$ dimensions.
The values for $\nu$, $\gamma$, $\theta_0$, $\theta_1$, $\theta_2$ are well established
for monodisperse chains \cite{DegennesBook,Duplantier86a,Duplantier89,DescloizBook}.
Using a coarse-grained computational model system of annealed polydisperse equilibrium polymers
(EPs) in strictly $d=2$ dimensions we verify here the exponents marked in bold.
The exponents $\delta$ and $\alpha$ in the last two lines characterize the average chain length 
$\avN \propto \exp(\delta E) \phi^{\alpha}$ with $\phi$ being the density and $E$ the scission energy 
of the EPs.
} 
\label{tab_expo}
\end{table}

\begin{figure}[t]
\centerline{\resizebox{0.9\columnwidth}{!}{\includegraphics*{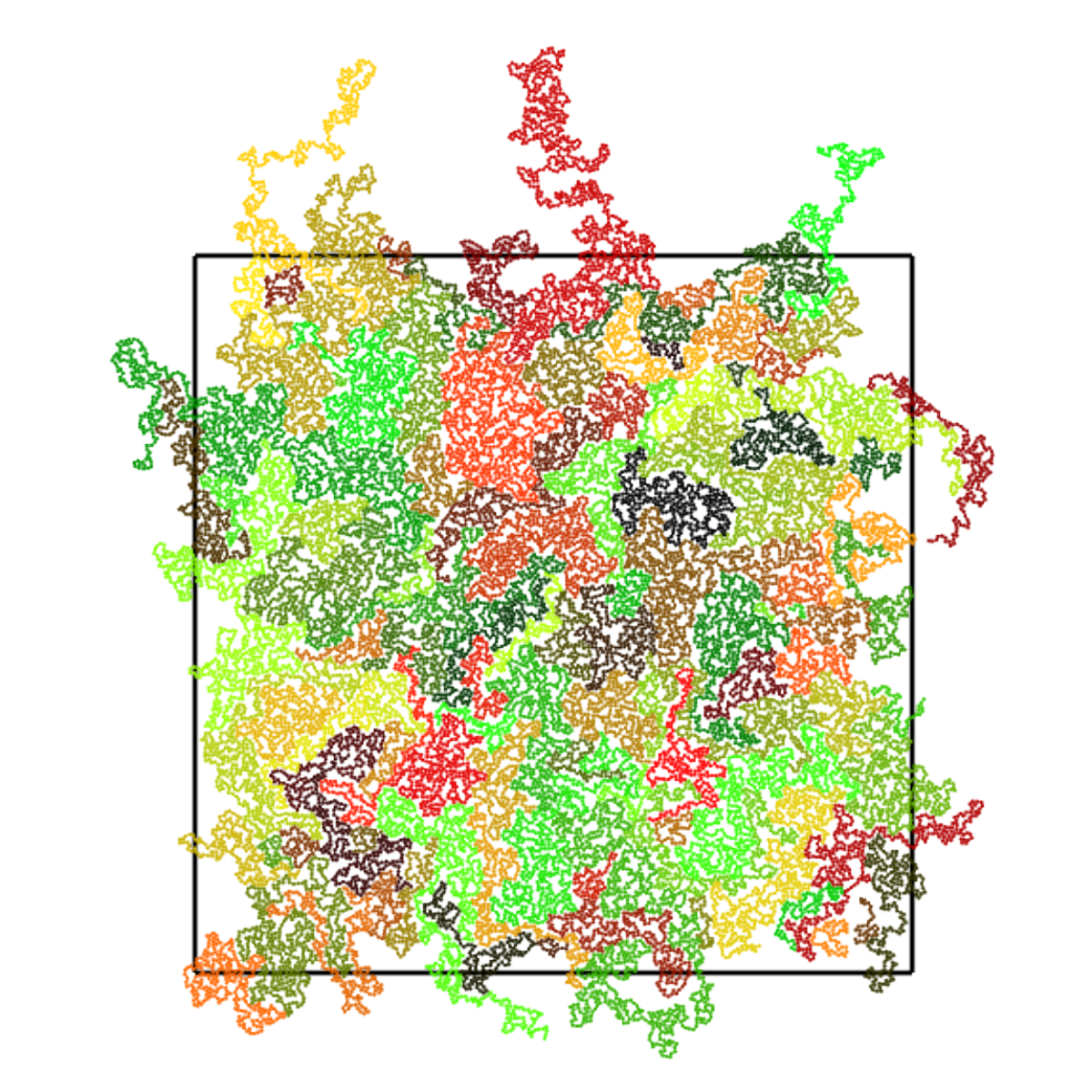}}}
\vspace*{-0.4cm}
\caption{Snapshot of EPs in $d=2$ without branching or ring formation
and disallowing any monomer overlap or chain intersection for a surface fraction 
$\phi = 0.39269$ of monomers and a scission energy $E=10$. 
The total configuration of $n=131072$ monomers is contained in a periodic square
simulation box of linear dimension $L=512$. Only a subvolume of $n=23060$ monomers
is shown representing all chains with at least one end in the square.
The empty white space is occupied by chains with ends outside.
}
\label{fig_snap}
\end{figure}

%
These observables are investigated here for different density limits 
by means of off-lattice Monte Carlo (MC) simulations \cite{AllenTildesleyBook,LandauBinderBook} 
of a simple coarse-grained model of annealed ``equilibrium polymers" (EPs)
where the polymerization takes place under condition of chemical equilibrium between the polymers and 
their respective monomers 
\cite{CC90,Jaric85a,Jaric85b,Milchev93_Potts,Milchev95,Milchev96,WMC98a,WMC98b,MWL00b}.
Theoretically such EPs are most readily described using a grand-canonical formalism \cite{DescloizBook}
as we shall also do below in Sec.~\ref{FH} by means of a Flory-Huggins mean-field approximation.
However, especially for dense solutions EPs do not only compete for monomers but also for space,
which may imply strong spatial correlations for these two-dimensional systems, and,
moreover, our simulations are strictly canonical, i.e. 
the total particle number $n$ of each simulation box is kept constant, 
and only for sufficiently large boxes both ensembles must become equivalent \cite{Callen}.
A snapshot of one configuration obtained at a moderately large surface fraction of disks is shown in 
Fig.~\ref{fig_snap}. While some chains appear to be compact (filling space densely), many are clearly 
not and most of the chain shapes are definitely not disklike with smooth perimeters.
The general question investigated in this work is whether the asymptotic exponents of monodisperse chains, 
both in the dilute and the dense limit, remain relevant for such EPs. More specifically, we will 
address the scaling of the typical chain size $R$, 
of the chain length distribution $p(N)$ and its first moment $\avN$,
of various intrachain distance distributions $G_i(r)$ 
--- allowing us to measure the ``contact exponents" $\theta_i$ --- 
and finally the intramolecular structure factor $F(q)$.
This will be done using two natural operational parameters tuning $\avN$:
the surface fraction $\phi$ of monomers and the ``scission energy" $E$ for breaking a bond.

It will be shown
that despite the intrinsic polydispersity EPs are characterized by the {\em same} exponents 
as their monodisperse peers. They adopt indeed at high densities and large chain lengths 
{\em on average} compact configurations with $\df =2$ and $\dperi=5/4$. 
Moreover, the average chain size $\avN$ increases as
\begin{equation}
\avN \propto \exp(\delta E) \phi^{\alpha}
\label{eq_alpha_delta_def}
\end{equation}
with respect to $\phi$ and $E$ in the respective density regimes.
As indicated in Table~\ref{tab_expo}, the two growth exponents $\delta$ and $\alpha$ are given by
\begin{equation}
\deltaDIL \equiv \alphaDIL \equiv \frac{1}{1+\gammaDIL} = \frac{32}{75}
\label{eq_deltaDILalphaDIL}
\end{equation}
in the dilute limit (as marked by the subscript ``$\DIL$"), 
\begin{equation}
\deltaDEN \equiv \frac{1}{1+\gammaDEN} = \frac{16}{35} 
\label{eq_deltaDEN}
\end{equation}
for all dense systems (subscript ``$\DEN$") and
\begin{equation}
\alphaSEM \equiv \deltaDEN \left[ 1 + \frac{\gammaDIL-\gammaDEN}{d \nuDIL -1} \right] = \frac{3}{5}
\label{eq_alphaSEM}
\end{equation}
in the semidilute regime (subscript ``$\SEM$").
While Eq.~(\ref{eq_deltaDEN}) is valid for all sufficiently large densities,
this is not the case for Eq.~(\ref{eq_alphaSEM}) 
which only applies for large semidilute blobs \cite{DegennesBook,foot_blob}.
Interestingly, $\alpha$ is found to increase more strongly for larger surface fractions
and it was the main numerical challenge of the presented work to demonstrate that 
$\alpha=\alphaSEM$ for the semidilute regime holds for sufficiently small densities and large chains
\cite{foot_alphaD3,Cates88,Schaefer92,Schoot97,Schurtenberger93,Schurtenberger96}.
%

We demonstrate first in Sec.~\ref{FH} the above exponents $\delta$ and $\alpha$ using a Flory-Huggins 
free energy approximation. We outline then in Sec.~\ref{sec_model} 
our computational model and its operational parameters before we turn in Sec.~\ref{sec_simu} 
to our numerical results.
A summary and an outlook to future work may be found in Sec.~\ref{sec_conc}.

\section{Flory-Huggins approximation}
\label{FH}

%
The main departure from the conventional theory of polymer solutions is that for EPs 
only the total monomer number $n$ is conserved and, hence, the monomer number density
$\rho=n/V$ with $V$ being the constant volume of the system, rather than the density 
distribution $c(N)$ of chains $N$ which is an annealed quantity.
(The number density $\rho$ is trivially proportional 
to the surface fraction $\phi$ used elsewhere in this work to characterize the density.)
The density distribution $c(N)$ is normalized such that
\begin{equation}
\rho = \sum_N N c(N).
\label{eq_FH_cN2rho}
\end{equation}
It is often more convenient to use instead the normalized 
number distribution $p(N) = \avN c(N)/\rho$ for which 
\begin{equation}
\sum_N p(N) =1 \mbox{ and } \avN= \sum_N N p(N) 
\label{eq_FH_pN_def}
\end{equation}
holds in agreement with Eq.~(\ref{eq_FH_cN2rho}).
Within the Flory-Huggins mean-field approximation \cite{CC90,Milchev93_Potts,Milchev95,Milchev96,WMC98a,WMC98b} 
the grand potential density $\Omega$ of the system may be written 
\begin{equation}
\Omega[c(N)] = \sum_N c(N) \left[ \ln[c(N)] + \fend + \mu N \right]
\label{eq_FH_functional}
\end{equation}
where we choose the energy units so that $\kBT=1$ and where irrelevant factors
(such as the persistence length) have been omitted for clarity.
The first term is the entropy of mixing,
the second term the free energy contribution due to the chain ends
and the last term entails the usual Lagrange multiplier for the conserved monomer density,
cf.~Eq.~(\ref{eq_FH_cN2rho}).
Without loss of generality, we have suppressed in Eq.~(\ref{eq_FH_functional})
the part of the free energy linear in chain length.
The chain end free energy contribution $\fend$ does in general depend on $E$, $N$ and $\rho$.
It is assumed to not dependent on the lengths of neighboring chains.
 
%
On the simplest mean-field level $\fend=E + const$ is given by the scission energy
and an irrelevant constant. Minimizing with respect to $c(N)$ and paying attention
to Eq.~(\ref{eq_FH_cN2rho}) yields $c(N) \propto \exp(-E-\mu N)$.
Hence,
\begin{equation}
\avN  = 1/\mu \propto \sqrt{\rho e^E}.
\label{eq_FH_MF}
\end{equation}
A comparison with the definitions made in Eq.~(\ref{eq_alpha_delta_def}) confirms 
the exponents $\delta=\alpha=1/2$ stated in the last two lines 
of the first column of Table~\ref{tab_expo}.
This result is only a good approximation near the $\theta$-temperature.
It coincides with the standard law of mass action \cite{Callen}.

%
For dilute two-dimensional EP in good solvent $\fend$ must also become chain length dependent,
\begin{equation}
\fend=E-(\gamma-1) \ln(N) \mbox{ with } \gamma=\gammaDIL=43/32,
\label{eq_FH_fend_DIL}
\end{equation}
due to the well-known enhancement factor for the partition function \cite{DegennesBook}.
Minimization of the total free energy with respect to $c(N)$ now leads to the
exponents $\deltaDIL=\alphaDIL=32/75$ stated above in Eq.~(\ref{eq_deltaDILalphaDIL}).
The normalized distribution $p(N)$ is given by a Schulz-Zimm (or Gamma) distribution 
\cite{Schulz39,Zimm48} 
\begin{equation}
p(x) = \frac{\gamma^{\gamma}}{\Gamma(\gamma)} x^{\gamma-1} \exp(-\gamma x) 
\mbox{ with } x = N/\avN
\label{eq_px_Schulz}
\end{equation}
being the reduced chain length,
$\Gamma(...)$ denoting the standard Gamma function \cite{abramowitz}
and $\gamma=\gammaDIL$. 
Note that Eq.~(\ref{eq_px_Schulz}) reduces to $p(x)=\exp(-x)$ under mean-field conditions ($\gamma = 1$)
while for $\gamma > 1$ the distribution must become non-monotonic with a depletion region for $x \ll 1$.

%
For dense EPs $\fend$ must depend both on $N$ and $\rho$.
It is given in the semidilute limit by
\begin{equation}
\fend=E-(\gammaDIL-1) \ln(g)-(\gammaDEN-1) \ln(N/g) 
\label{eq_FH_fend_SEM}
\end{equation} 
for chain lengths $N \gg g$ with $g$ being the number of monomers in a semidilute blob in $d=2$.
The second term is needed to match Eq.~(\ref{eq_FH_fend_SEM}) at $N=g$ with the corresponding
dilute free end contribution Eq.~(\ref{eq_FH_fend_DIL}). The last term corresponds to the
enhancement factor of the partition function of a chain of blobs.
(This term is not present for dense EPs in $d=3$ for which $\gammaDEN=1$ \cite{CC90,WMC98a,WMC98b}.)
Minimizing again the Flory-Huggins approximation Eq.~(\ref{eq_FH_functional})
leads to the exponents $\delta=\deltaDEN=16/35$ and $\alpha=\alphaSEM=3/5$
stated in the Introduction, cf. Eq.~(\ref{eq_deltaDEN}) and Eq.~(\ref{eq_alphaSEM}).
It was used here that \cite{DegennesBook,foot_blob}
\begin{equation}
g(\rho) \approx \rho \xi^d(\rho) \propto 1/\rho^{\frac{1}{\nuDIL d -1}} = 1/\rho^2
\label{eq_FH_grho}
\end{equation}
with $\xi(\rho) \propto g(\rho)^{\nuDIL} \propto 1/\rho^{3/2}$ 
being the size (correlation length) of the semidilute blob.
More generally, $\fend$ is given in the dense limit by
\begin{equation}
\fend=E+\funknown(\rho)-(\gammaDEN-1) \ln(N) 
\label{eq_FH_fend_MEL}
\end{equation}
with $\funknown(\rho)$ being an apriori unknown function. Hence,
\begin{equation}
\avN \propto\left[  e^{ E} \rho e^{\funknown(\rho)} \right]^{\deltaDEN}.
\label{eq_FH_N_MEL}
\end{equation}
In this case $\delta=\deltaDEN$ still holds while the (effective) growth exponent $\alpha$ must be fitted. 
Let us assume that $\funknown(\rho) = \aunknown \ln(\rho)$ increases logarithmically 
with an unknown coefficient $\aunknown$. It follows then from Eq.~(\ref{eq_FH_N_MEL}) that 
\begin{equation}
\alpha = \deltaDEN ( 1 + \aunknown).
\label{eq_FH_aunknown2alpha}
\end{equation}
We note finally that quite generally the normalized distribution $p(N)$ in the dense regime is 
still given by the Schulz-Zimm distribution Eq.~(\ref{eq_px_Schulz}), however, now with an 
exponent $\gamma=\gammaDEN=19/16$.

%
We have assumed above that the Flory-Huggins approximation Eq.~(\ref{eq_FH_functional})
holds for all densities. While this is trivial in the dilute 
limit (where chains barely interact) this must be checked for dense systems 
since the lengths of neighboring chains may be correlated. 
This is where simulations of simple model systems become important.

\section{Computational model}
\label{sec_model}

%
At variance to permanent polymer chains commonly considered in computer simulations 
\cite{VanderzandeBook,LandauBinderBook,CarmesinKremer90} EPs have a finite scission energy $E$ 
attributed to each bond.
Following previous work \cite{WMC98a,WMC98b,WJC10} $E$ is assumed to be independent of density,
chain length and the curvilinear position of the bond. 
It has to be paid whenever a bond between two monomers is broken.
Specifically, we investigate here by means of off-lattice MC simulations \cite{AllenTildesleyBook}
a simple polymer model of monodisperse hard disks of diameter $\sigma$
where each disk may be bonded to at most two other disks (no branching).
By restricting the distance $r$ between bonded disks to $r < 1.4 \sigma$ the intersection 
of two chains of disks becomes impossible as may be seen from Fig.~\ref{fig_snap}.
The bonds between disks are reversibly broken and recombined by means of a Metropolis scheme 
\cite{AllenTildesleyBook,WMC98b}.
Only linear chains are present and the formation of closed rings is explicitly forbidden.
We use reduced units with $T=\kB=\sigma=1$ for, respectively,
temperature, Boltzmann's constant and disk diameter.

\begin{figure}[t]
\centerline{\resizebox{0.9\columnwidth}{!}{\includegraphics*{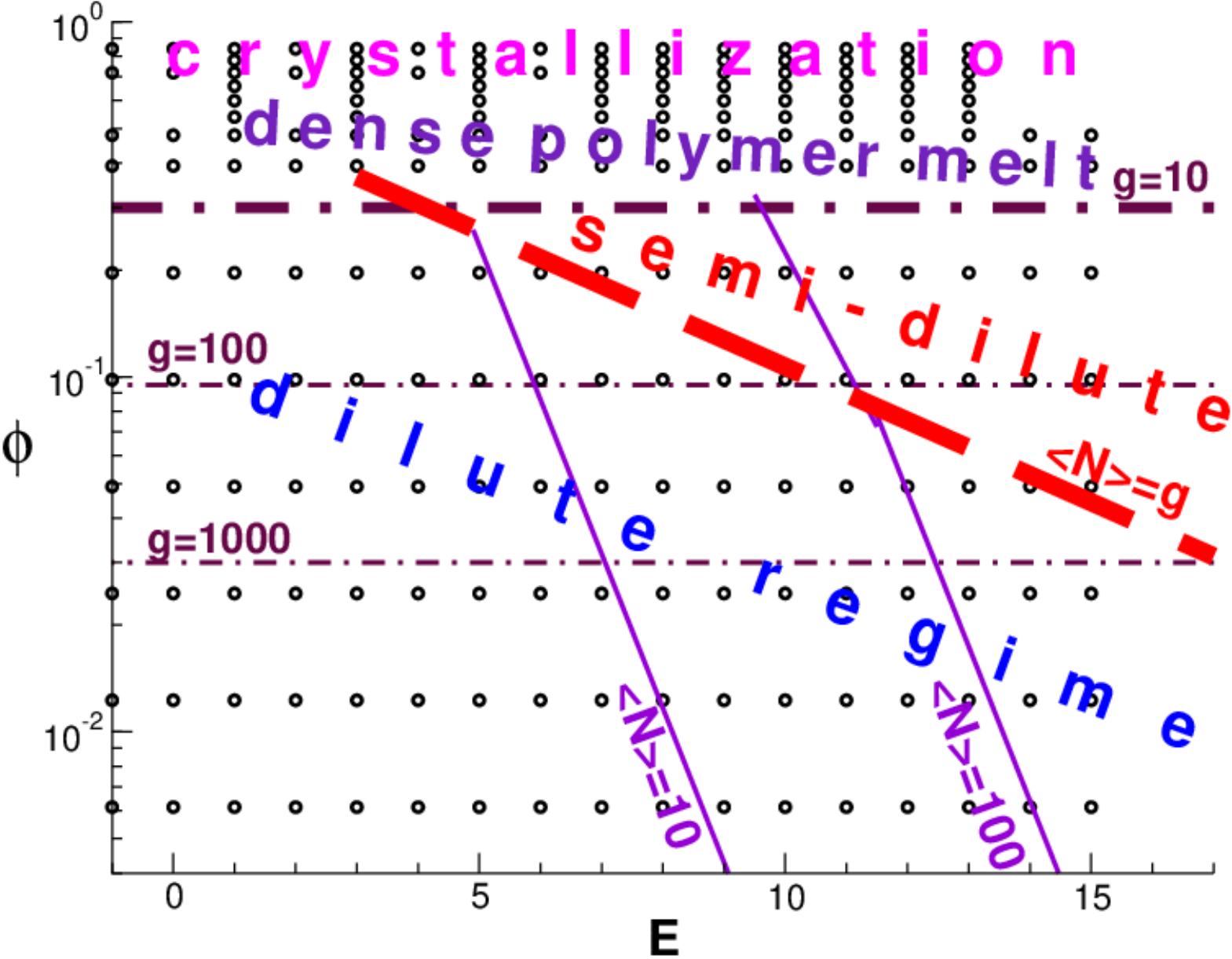}}}
\caption{Operational parameters $(E,\phi)$ simulated (small circles)
and sketch of different regimes discussed below.
The systems below the bold dashed line are dilute, i.e.
the typical chain size $\avN$ is much smaller than the number 
$g(\phi) \approx \phi \xi(\phi)^d \propto 1/\phi^2$
of monomers contained in a semidilute blob of correlation length 
$\xi(\phi) \propto 1/\phi^{3/2}$ for asymptotically large chains \cite{DegennesBook}. 
Semidilute behavior with $\avN \propto \phi^{\alpha}$ and a growth-exponent $\alpha = 3/5$ 
is found for systems above the dashed line and below the horizontal (dash-dotted) line for $g=10$. 
}
\label{fig_phiE}
\end{figure}
 
Since the bonds between monomers constantly break and recombine it is inefficient
to base the data structure on the chains which would be penalized by either large sorting times
or a waste of computer memory. Instead one has to base the data structure on the bonds of the
monomers using a pointer list between connected bonds. 
As described elsewhere\cite{WMC98b}
this avoids all sorting procedures at the expense of one additional pointer list of length $2n+1$. 
In order to obtain configurational properties, such as the radius of gyration,
one must unfortunately either finally run a sorting routine to bring the data into a conventional form
suitable for the standard routines or, as we have done, rewrite all needed analysis tools
in terms of the pointer list connecting the monomer bonds.

\begin{table}[t]
\begin{tabular}{|c|c|c|c||c|c|c|c|c|}
\hline
$\phi$   & $L$  & $n$    & $E$   & $\avM$& $\avN$& $l$    & $\Ren$ & $\Rgn$\\  \hline
0.838896 & 256  & 70000  & 10    & 160   &  438  & 1.041  & 29     & 13    \\
0.838896 & 256  & 70000  & 13    & 44    &  1584 & 1.041  & 53     & 23    \\
0.479369 & 256  & 40000  & 10    & 149   &  269  & 1.194  & 30     & 13    \\
0.479369 & 256  & 40000  & 15    & 17    &  2401 & 1.195  & 88     & 38    \\
0.392699 & 512  & 131072 & 10    & 578   &  227  & 1.206  & 30     & 13    \\
0.392699 & 256  & 32768  & 15    & 15    &  2205 & 1.206  & 89     & 42    \\
0.392699 & 512  & 131072 & 15    & 66    &  1981 & 1.206  & 91     & 38    \\
0.098175 & 1024 & 131072 & 10    & 1752  &  75   & 1.219  & 28     & 11    \\
0.098175 & 1024 & 131072 & 15    & 175   &  748  & 1.219  & 107    & 45    \\
0.024544 & 2048 & 131072 & 10    & 3917  &  34   & 1.220  & 17     & 6     \\
0.024544 & 2048 & 131072 & 15    & 455   &  288  & 1.220  & 84     & 3     \\
0.006135 & 2048 & 32768  & 15    & 230   &  143  & 1.220  & 53     & 19    \\ 
0.001533 & 4096 & 32768  & 15    & 421   &  78   & 1.220  & 33     & 17    \\
\hline
\end{tabular}
\caption[]{Selection of operational parameters $\phi$, $L$, $n$ and $E$ used
and of some properties obtained:
typical chain number $\avM=n/\avN$ per simulation box,
typical chain length $\avN$, 
typical root-mean squared bond length $l$,
number-averaged end-to-end distance $\Ren$ and gyration radius $\Rgn$.
} 
\label{tab_data}
\end{table}

%
As summarized in Fig.~\ref{fig_phiE} we vary $E$ between $-1$ and $15$ and the 
monomer surface fraction $\phi$ by several orders of magnitude. 
Periodic square simulation boxes of linear size $L$ have been used.
We have checked for the finite system size effects expected for the largest $E$ \cite{WMC98b}
and it was occasionally found necessary to further increase the box sizes.
The number of particles $n \propto \phi L^2$ ranges between $2^{15}$ and $2^{18}$.
Table~\ref{tab_data} presents a small selection of operational parameters used
and of some of the properties obtained. Naturally, the number of chains $\avM=n/\avN$ per simulation box
must be sufficiently large to avoid finite box-size effects. Systems with large $E$,
such as the example for $\phi=0.479369$ and $E=15$ with $\avM \approx 17$ 
(cf.~Table~\ref{tab_data}),
have to be considered with care.
 
%
Only local MC hopping moves are needed for sampling the configuration space
since the breaking and recombination of chains reduces the relaxation times dramatically 
if large scission-recombination frequencies are used \cite{WMC98b}.
%
We run each equilibrated configuration over $10^8$ MC steps 
while storing $1000$ ``frames" at equidistant time intervals. 
The static properties of these frames are then analysed and finally averaged. 
%
As one expects, the monodisperse disks form essentially hexagonal packings
for our largest $\phi$ \cite{foot_phihigh,Zaccone22}. (This can be simply seen
by inspection of snapshots or by computing the usual total structure factor.)
Interestingly, albeit the monomers barely move in this limit the bonds between monomers 
still rearrange and the chain connectivity remains an annealed property. 
We are thus able to equilibrate chain properties up to $E=13$.
This would have been impossible for monodisperse chains at similar densities.

\section{Numerical results}
\label{sec_simu}

\subsection{Typical chain size $R$}
\label{sub_RN}

\begin{figure}[t]
\centerline{\resizebox{0.9\columnwidth}{!}{\includegraphics*{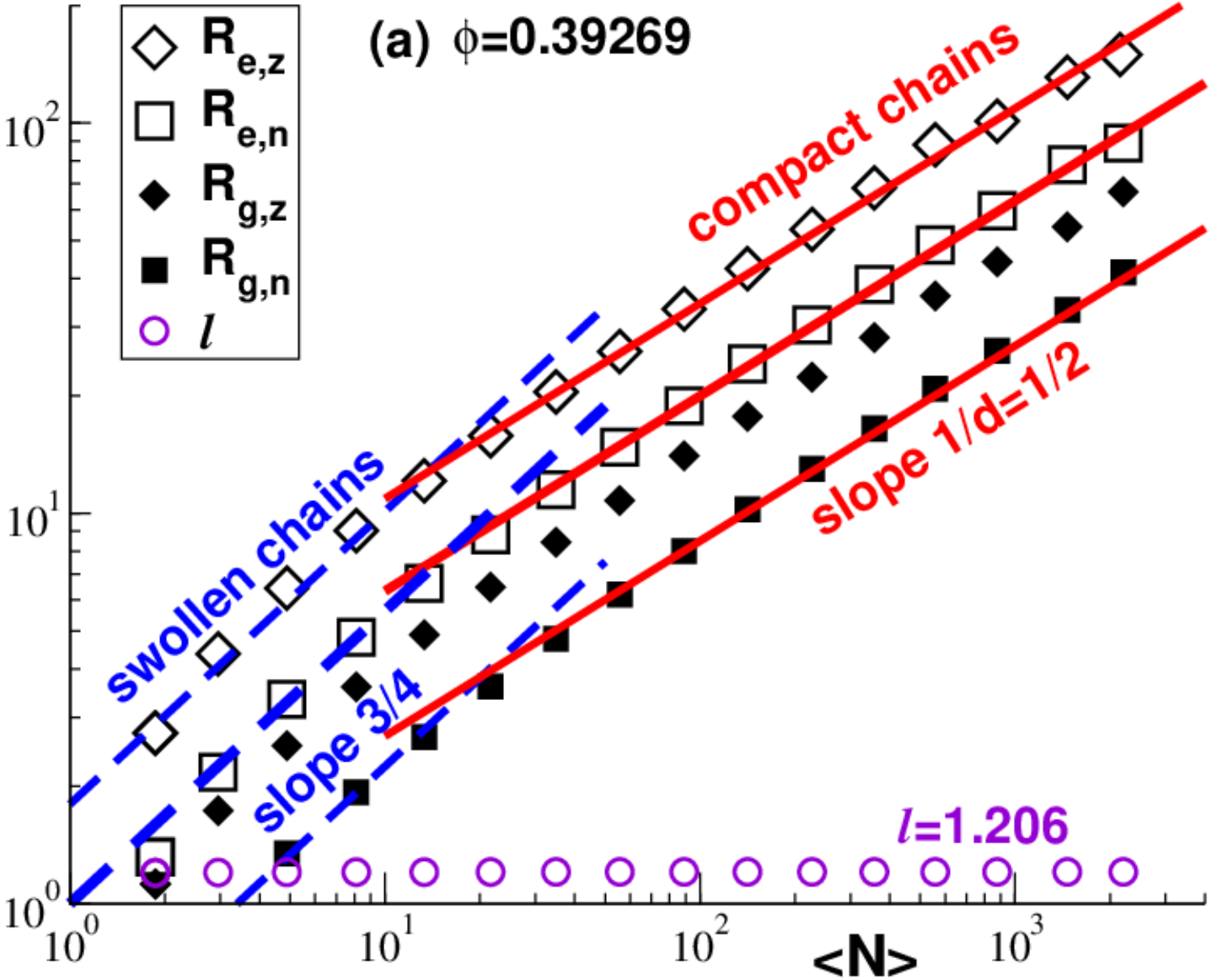}}}
\vspace*{0.15cm}
\centerline{\resizebox{0.9\columnwidth}{!}{\includegraphics*{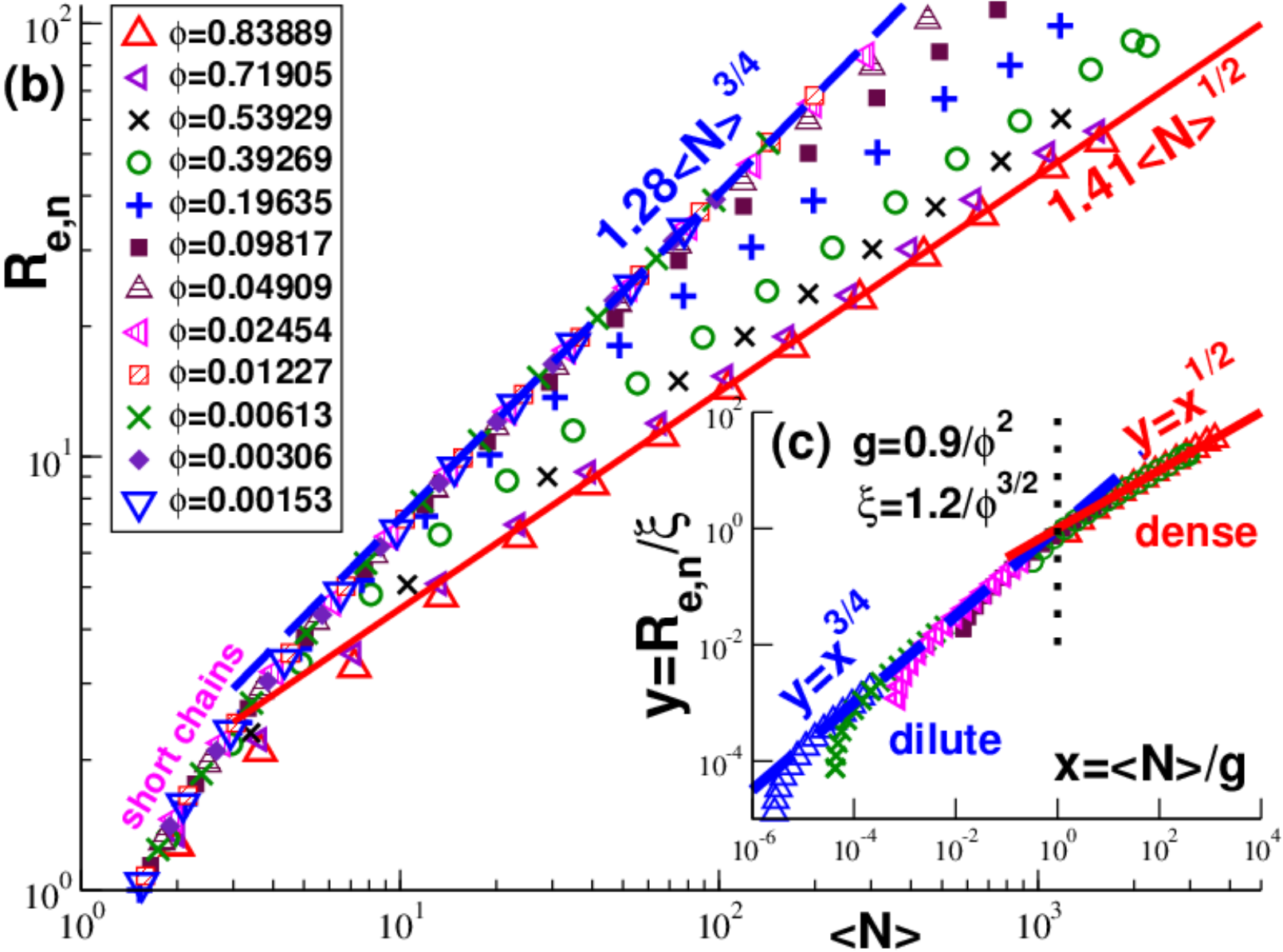}}}
\caption{Typical chain size $R$ as a function of $\avN$ 
with dashed lines indicating the Flory exponent $\nuDIL=3/4$ and bold lines $\nuDEN=1/2$:
{\bf (a)} 
$n$- and $z$-averages $\Ren$ and $\Rez$ for the end-to-end distance (open symbols) and 
$\Rgn$ and $\Rgz$ for the radius of gyration (filled symbols) for $\phi=0.39269$. 
Compact behavior is found for $\avN \gg 10$. 
The typical bond length $l$ (circles) is essentially constant.
{\bf (b)}
End-to-end distance $\Ren$ for a broad range of $\phi$
showing that the chains become compact with increasing $\phi$ and $\avN$.
{\bf (c)}
Successful density crossover scaling for $\Ren/\xi(\phi)$ {\em vs.} $\avN/g(\phi)$ 
for a large range of densities using $\xi = 1.2/\phi^{3/2}$ and $g=0.9/\rho^{2}$. 
}
\label{fig_RN}
\end{figure}

The typical size $R$ for EPs of different $(E,\phi)$ is characterized
in Fig.~\ref{fig_RN} as a function of the mean chain length $\avN$.
All the presented $R$ are root-mean-square averages,
i.e. the square root is taken in a final step after the averaging procedure.
We compute in a first step the typical squared end-to-end distance $\Rend^2(N)$ and
radius of gyration $\Rgyr^2(N)$ for each chain length $N$ and then in a second step
the ``$n$-averages" $\Ren^2$ and $\Rgn^2$ over the normalized number distribution $p(N)$ and
the ``$z$-averages" $\Rez^2$ and $\Rgz^2$ with a weight proportional to $N^2 p(N)$
\cite{DescloizBook,foot_averages}.
(We remind that experimentally the $z$-average $\Rgz$ is relevant for neutron scattering \cite{BenoitBook}.)

Data obtained for $\phi=0.39269$ is presented in panel {\bf (a)} of Fig.~\ref{fig_RN}.
(Also included is the typical bond length $l$.)
Albeit larger $N$ have a stronger weight for the $z$-average all
presented data reveal the same scaling behavior:
while short chains show a power-law exponent $\nu=3/4$
(dashed line) consistent with swollen chain statistics in the dilute limit
larger chains are clearly compact ($\nu=1/d$).
A broad range of densities is presented in panel {\bf (b)} where we trace $\Ren$ as a function of $\avN$.
As can be seen, the data becomes $\phi$-independent in the low-$\phi$ limit
where $\Ren \simeq 1.28 \avN^{3/4}$ as emphasized by the dashed bold line. 
This holds for sufficiently large $\avN \gg 10$. Obviously, $\Ren$ and $\avN$ may not
become too large for a given density, i.e. $\Ren \ll \xi(\phi)$ and $\avN \ll g(\phi)$
must hold for a given $\phi$. 
As may be seen for $\phi=0.09817$,
the effective power-law slope may eventually approach $\nu = 1/2$ in the opposite limit. 
As shown by the data for $\phi=0.83889$ (bold solid line) the dilute $\avN$-regime 
with $\nu=3/4$ eventually becomes irrelevant for the largest surface fractions
where the chains are compact for $\avN \gg 10$.
A classical density crossover scaling following De~Gennes \cite{DegennesBook}
is presented in Fig.~\ref{fig_RN}(c) where we focus again on $\Ren$. 
Using Eq.~(\ref{eq_FH_grho}) 
we trace the rescaled chain size $y=\Ren/\xi$ as a function of the rescaled chain length $x=\avN/g$.
A successful data collapse for a broad range of densities is observed.
The prefactors for $\xi(\phi)$ and $g(\phi)$ are determined by imposing the matching 
of the asymptotic low- and high-density slopes at $(x,y)=(1,1)$ \cite{foot_blob}.
The existing deviations from the dilute-semidilute scaling for large $\phi$
(above the upper horizontal line in Fig.~\ref{fig_phiE}) are surprisingly 
small in this logarithmic representation.

\subsection{Typical chain length $\avN$}
\label{sub_N}
\begin{figure}[t]
\centerline{\resizebox{0.9\columnwidth}{!}{\includegraphics*{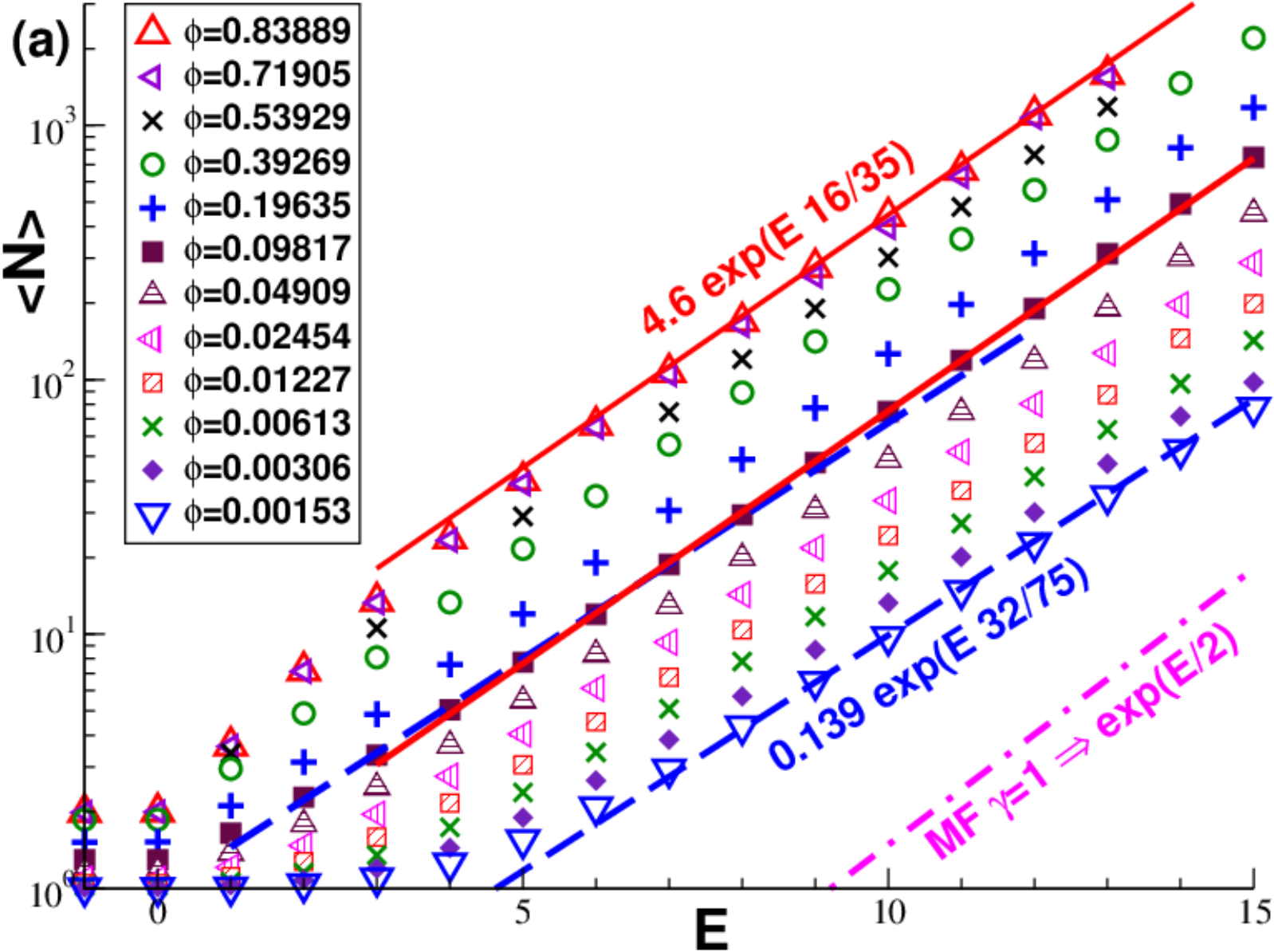}}}
\centerline{\resizebox{0.9\columnwidth}{!}{\includegraphics*{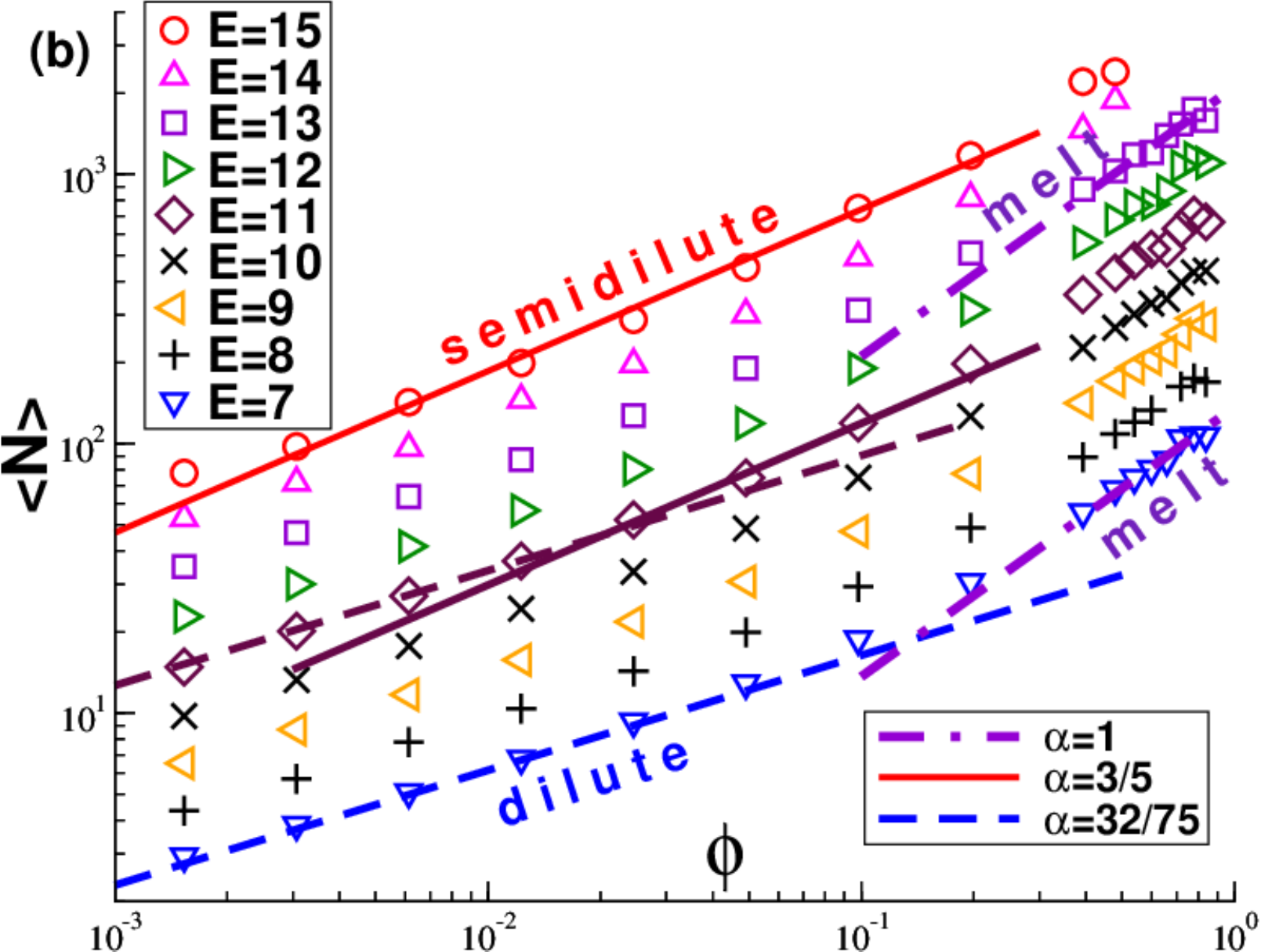}}}
\caption{Unscaled typical chain size $\avN$ as a function of 
{\bf (a)} scission energy $E$ for different $\phi$ (half-logarithmic representation) and
{\bf (b)} surface fraction $\phi$ for different $E$ (double-logarithmic representation).
The dashed (solid) lines indicate the expected behavior for dilute (dense) systems.
Note that the predicted growth exponent $\alpha =3/5$ for semidilute solutions in panel {\bf (b)}
can only be observed for the largest $E$ and not too large $\phi$ while a stronger 
$\phi$-increase is seen for $\phi > \phistarstar \approx 0.2$. 
The dashed-dotted lines indicate for this
``melt regime" an empirical power-law exponent $\alpha \approx 1$.
}
\label{fig_N}
\end{figure}

\begin{figure}[t]
\centerline{\resizebox{0.9\columnwidth}{!}{\includegraphics*{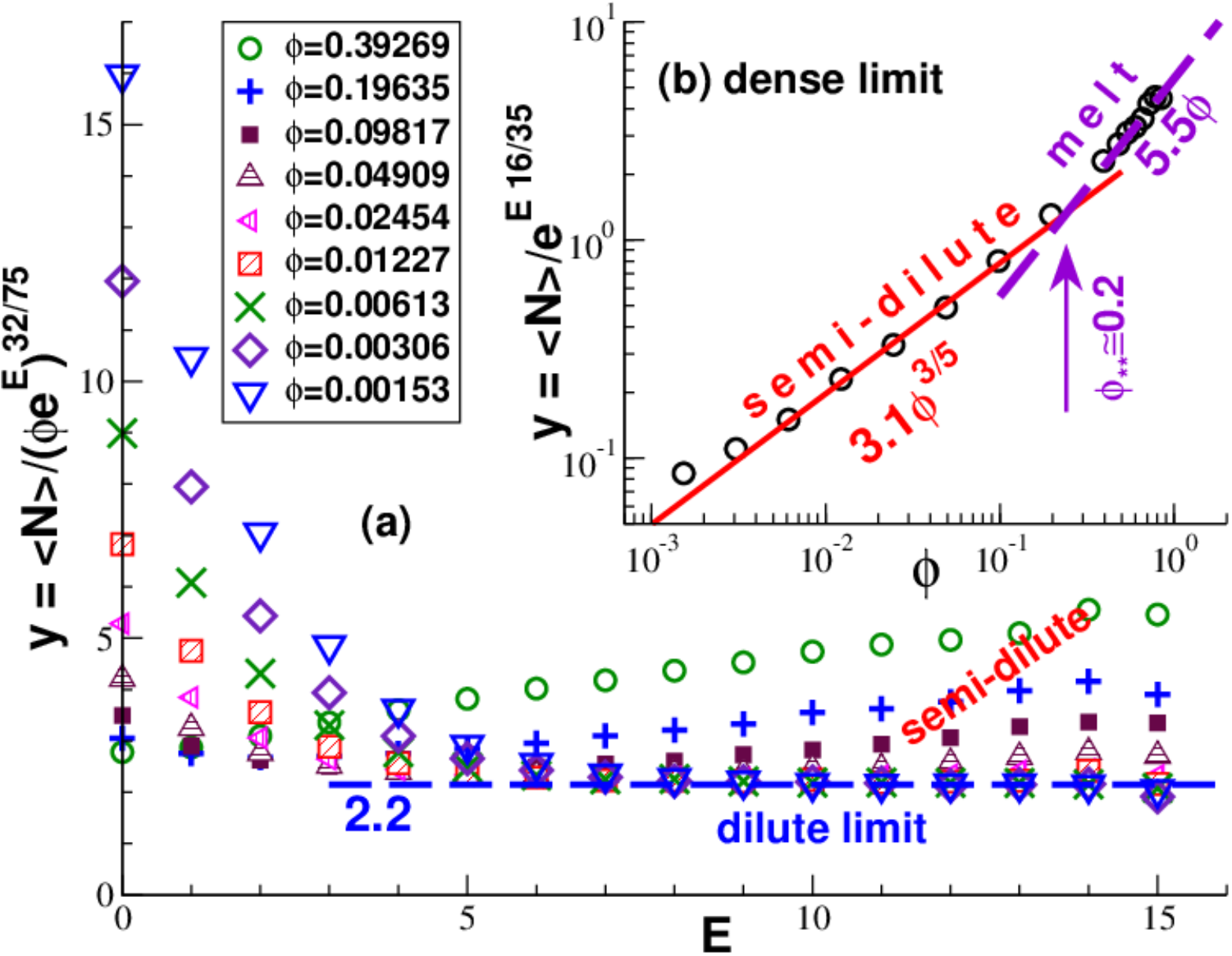}}}
\centerline{\resizebox{0.9\columnwidth}{!}{\includegraphics*{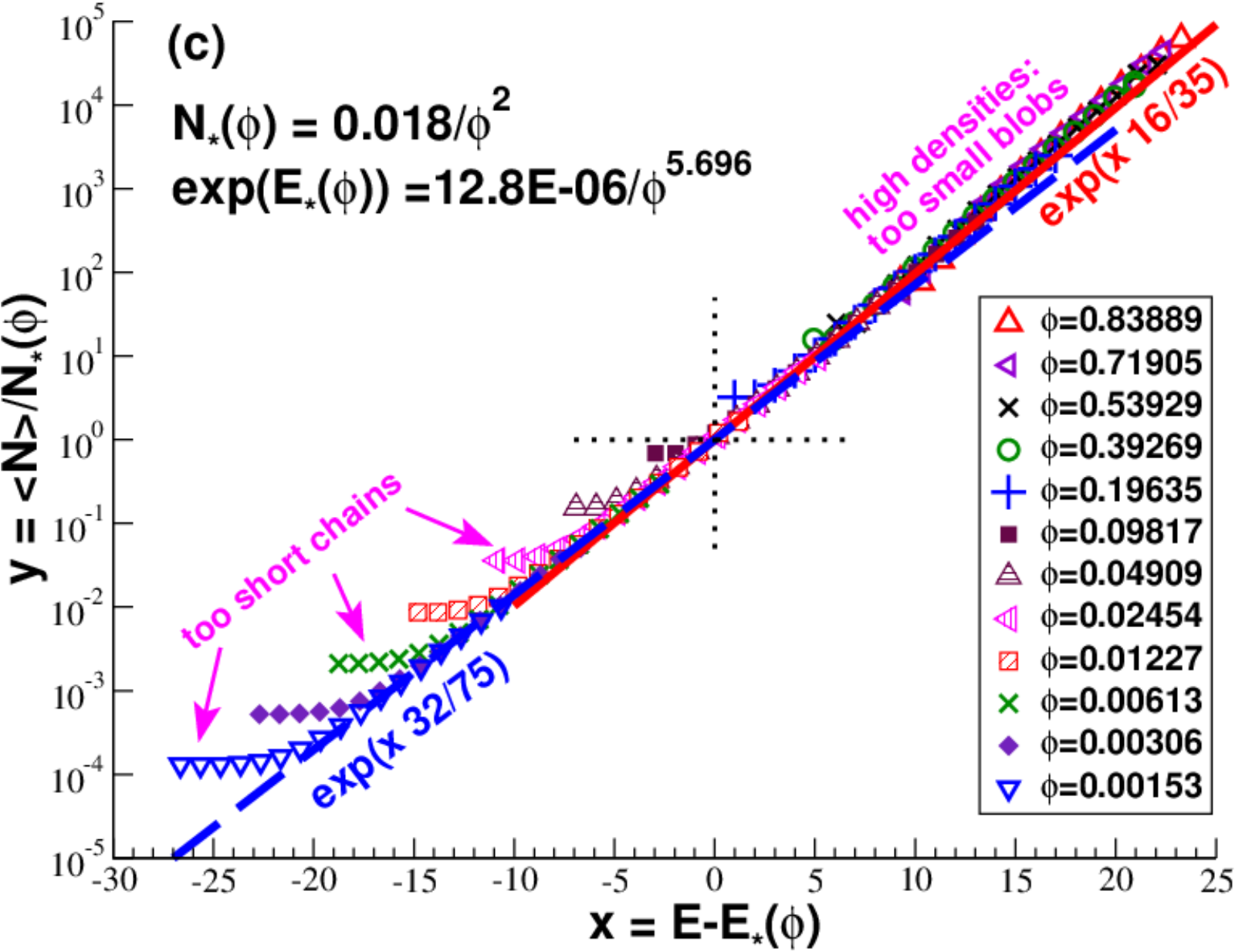}}}
\caption{Crossover scaling for $\avN$:
{\bf (a)} Linear representation of
$y=\avN/(\phi e^E)^{32/75}$ {\em vs.} $E$ 
used for a careful verification of the dilute reference (dashed line),
{\bf (b)}
double logarithmic representation of $y=\avN/e^{E 16/35}$ {\em vs.} $\phi$
for the largest available $E$ demonstrating $\alpha = 3/5$ ($\alpha=1$)
for the semidilute (melt) regime and
{\bf (c)}
successful dilute-semidilute crossover scaling $\avN/\Nstar(\phi)$ {\em vs.} 
$x=E-\Estar(\phi)$ using a half-logarithmic representation.
Deviations are seen in the dilute limit ($x \ll 0$) for too short chains and for 
$x \gg 10$ corresponding to systems in the melt regime ($\phi > \phistarstar \approx 0.2$).
}
\label{fig_Nscal}
\end{figure}

We investigate now in Fig.~\ref{fig_N} and Fig.~\ref{fig_Nscal}
the average chain length $\avN$ comparing our numerical results
with the exponents stated in the Introduction.
%
We present in Fig.~\ref{fig_N} the unscaled $\avN$ either
in panel {\bf (a)} as a function of the scission energy $E$ for a broad range of surface fractions $\phi$ 
or in panel {\bf (b)} as a function of $\phi$ for different $E$.
The dashed lines indicate the power-law slopes expected in the dilute limit,
the solid lines the corresponding ones for the dense limit.
Importantly, the exponent $\deltaDEN=16/35$ is seen in panel {\bf (a)}
to perfectly hold up to the highest densities. 
This is different for the exponent $\alpha$ presented in panel {\bf (b)}.
In fact, the exponent $\alpha=\alphaSEM=3/5$ for the semidilute
regime only holds for our largest scission energies $E$
and for surface fractions $\phi \ll \phistarstar \approx 0.2$.
Interestingly, $\alpha$ increases more strongly for denser systems
where an apparent exponent $\alpha \approx 1$ (dashed-dotted lines) is fitted.
Not surprisingly, the semidilute density dependence assumed in Eq.~(\ref{eq_FH_fend_SEM})
and Eq.~(\ref{eq_FH_grho}) becomes inappropriate in this limit and must be replaced by a 
more general density dependence as discussed above, cf.~Eq.~(\ref{eq_FH_N_MEL}). 
Using Eq.~(\ref{eq_FH_aunknown2alpha}) a measured apparent $\alpha \approx 1$
implies $\aunknown \approx 1$ for the phenomenological coefficient of an assumed
logarithmic density contribution to the free energy $\fend$.
(A value $\alpha \approx 1$ was also found for an EP bead-spring model in $d=3$ \cite{MWL00b}.)

The dilute-semidilute density crossover is further characterized in Fig.~\ref{fig_Nscal}.
To test the behavior in the dilute limit we have traced in panel {\bf (a)}
the reduced chain length $y=\avN/(\phi e^E)^{32/75}$ as a function of $E$.
For sufficiently small $\phi$ and large $\avN$ all data
collapse onto $y \approx 2.2$ (horizontal dashed line). 
Focusing on the largest $E$ available for all $\phi$ we trace in panel {\bf (b)}
the reduced chain length $y=\avN/e^{E 16/35}$ as a function of $\phi$.
We have thus taken as reference the expected $E$-dependence in the dense limit.
As emphasized by the bold solid line, the semidilute exponent $\alpha=\alphaSEM=3/5$
holds over about an order of magnitude.
A successful dilute-semidilute crossover scaling is presented in panel {\bf (c)}.
We trace here the reduced average chain length $y=\avN/\Nstar(\phi)$ as a function
of the shifted scission energy $x=E-\Estar(\phi)$ for a broad range of surface fractions.
Note that $\Nstar(\phi)$ is similar to $g(\phi)$, just with a slightly smaller prefactor. 
The shift scission energy $\Estar(\phi)$
is obtained from matching considerations which lead to
\begin{equation}
\exp(\Estar(\phi)) = (3.1/2.2)^{\frac{1}{\deltaDIL-\deltaDEN}}/\phi^{\frac{\alphaDEN-\alphaDIL}{\deltaDEN-\deltaDIL}}
\label{eq_Estar}
\end{equation}
with $2.2$ being the amplitude obtained in panel {\bf (a)}
and $3.1$ the amplitude determined in panel {\bf (b)} of Fig.~\ref{fig_Nscal}.
Deviations from the assumed scaling are seen for $x \gg 10$. 
These data points stem from dense systems above the upper horizontal line in Fig.~\ref{fig_phiE}. 

\subsection{Chain length number distribution $p(N)$}
\label{sub_cN}

\begin{figure}[t]
\centerline{\resizebox{0.9\columnwidth}{!}{\includegraphics*{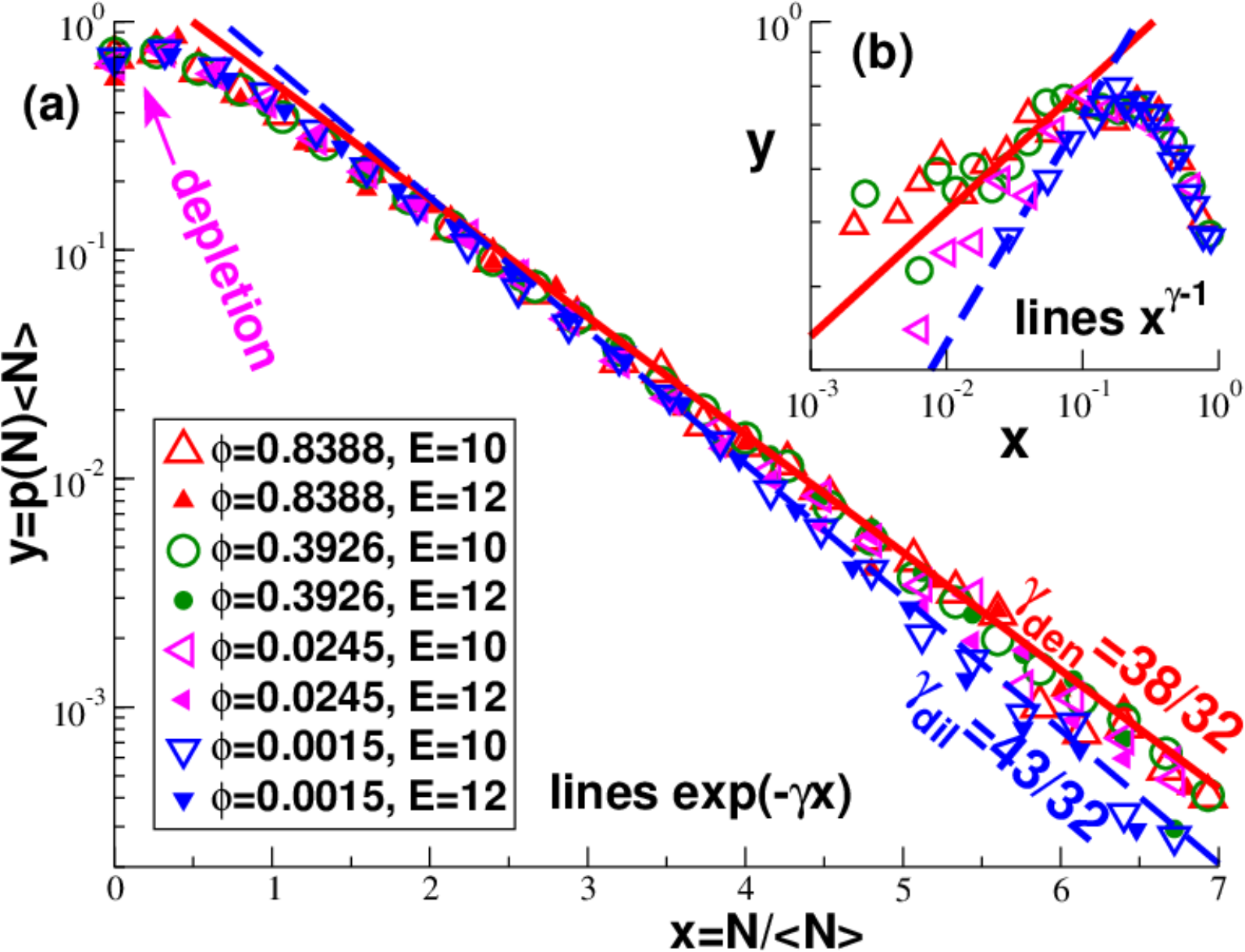}}}
\caption{Rescaled normalized number distribution $y=p(N) \avN$ {\em vs.} 
reduced chain length $x=N/\avN$ for different surface fractions $\phi$ and
scission energies $E$ as indicated in the legend:
{\bf (a)} 
Half-logarithmic representation focusing on large $x$
where $y \simeq \exp(-\gamma x)$ is expected,
{\bf (b)}
double-logarithmic representation focusing on the depletion limit for $x \ll 1$
where $y \propto x^{\gamma-1}$ is expected.
The dashed lines correspond to the exponent $\gammaDIL=43/32$ for the dilute limit,
the bold solid lines to $\gammaDEN=38/32$ for dense EP.
}
\label{fig_cN}
\end{figure}

The average chain length $\avN$ is the first moment of the normalized chain length
distribution $p(N)$ which contains {\em apriori} more information but is more difficult to measure.
We present in Fig.~\ref{fig_cN} $p(x) = p(N) dN/dx$ with $x = N/\avN$ being the reduced chain length.
As pointed out in Sec.~\ref{FH} a Schulz-Zimm distribution, cf.~Eq.~(\ref{eq_px_Schulz}), 
is expected in both density limits characterized by the respective values of the exponent $\gamma$: 
$\gammaDIL=43/32$ and $\gammaDEN=19/16$. 
Despite the small difference between both exponents 
the data presented in Fig.~\ref{fig_cN} for a broad range of densities
is clearly consistent with both the general form Eq.~(\ref{eq_px_Schulz}) 
of the distribution and the exponents in the respective density limits.
Interestingly, the asymptotic limit $p(x) \simeq \exp( -\gammaDEN x)$ for $x \gg 1$ indicated by the 
bold solid line in panel {\bf (a)} holds even up to very high surface fractions where the disks crystallize.
Please note that we have indicated data for scission energies only up $E=12$ since 
for higher values the statistics deteriorates. 
(To check the respective exponents for larger $E$ it is better to check the scaling of $\avN$.)
It is important to emphasize that all sampled $p(x)$ decay exponentially, 
at least for sufficiently large system sizes. This implies that all standard moments and variances 
do exist and {\em are not dominated by properties
which may be relevant in the tails of the distributions}.
 
As shown in panel {\bf (b)} of Fig.~\ref{fig_cN} the chain length distribution
reveals clearly a depletion for $x \ll 1$. The dashed and the solid slopes
indicate the power law $x^{\gamma-1}$ expected for, respectively,
dilute and dense systems. The accuracy of our data is, obviously,
insufficient to unambiguously demonstrate these power laws.
However, it is clear that there is no evidence in our data for
a {\em negative} exponent as postulated in some treatments of the unusual
diffusive behavior in dense three-dimensional giant micelles \cite{Langevin90}.

\subsection{Segment size distributions $G_i(r,s)$}
\label{sub_Gi}

\begin{figure}[t]
\centerline{\resizebox{0.9\columnwidth}{!}{\includegraphics*{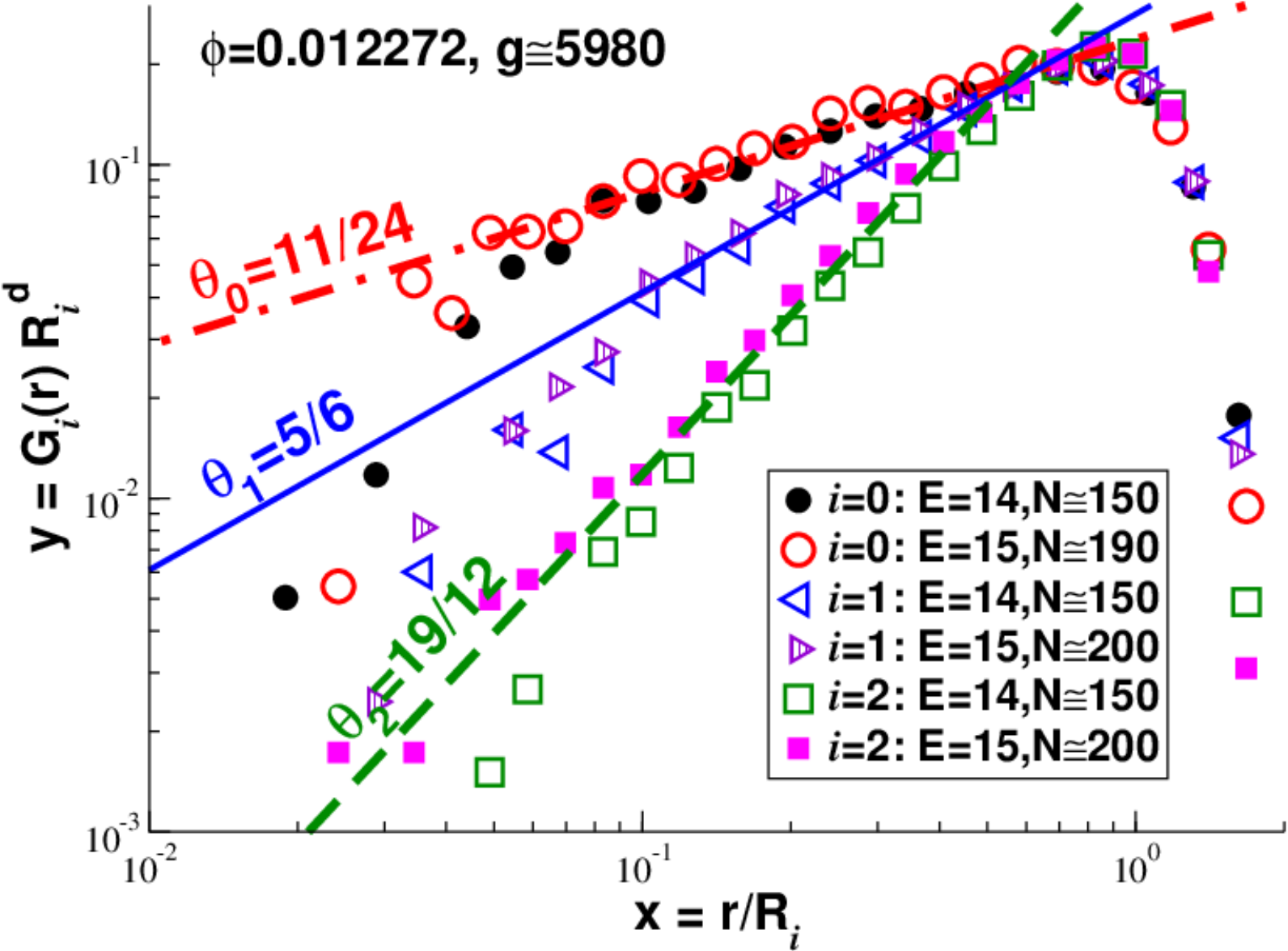}}}
\caption{Rescaled size distributions $y= G_i(r) R_i^2$ {\em vs.} $x=r/R_i$
for $i=0, 1$ and $2$ focusing on the dilute density limit with $\phi=0.012262$ 
for different $E$ and $N$-windows as given in the legends.
The slopes indicate the expected exponents.
}
\label{fig_Gi_dilute}
\end{figure}

\begin{figure}[t]
\centerline{\resizebox{0.9\columnwidth}{!}{\includegraphics*{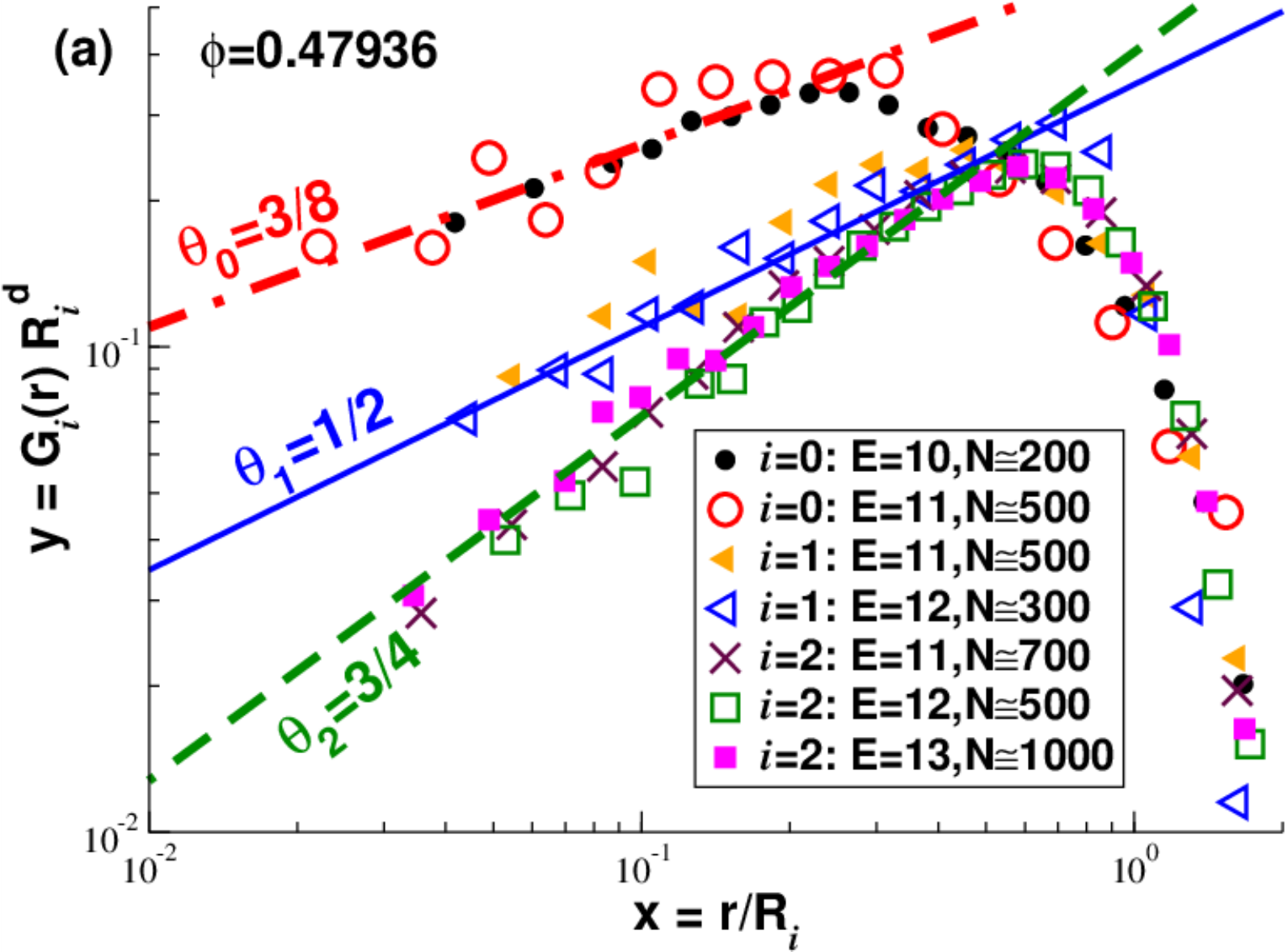}}}
\centerline{\resizebox{0.9\columnwidth}{!}{\includegraphics*{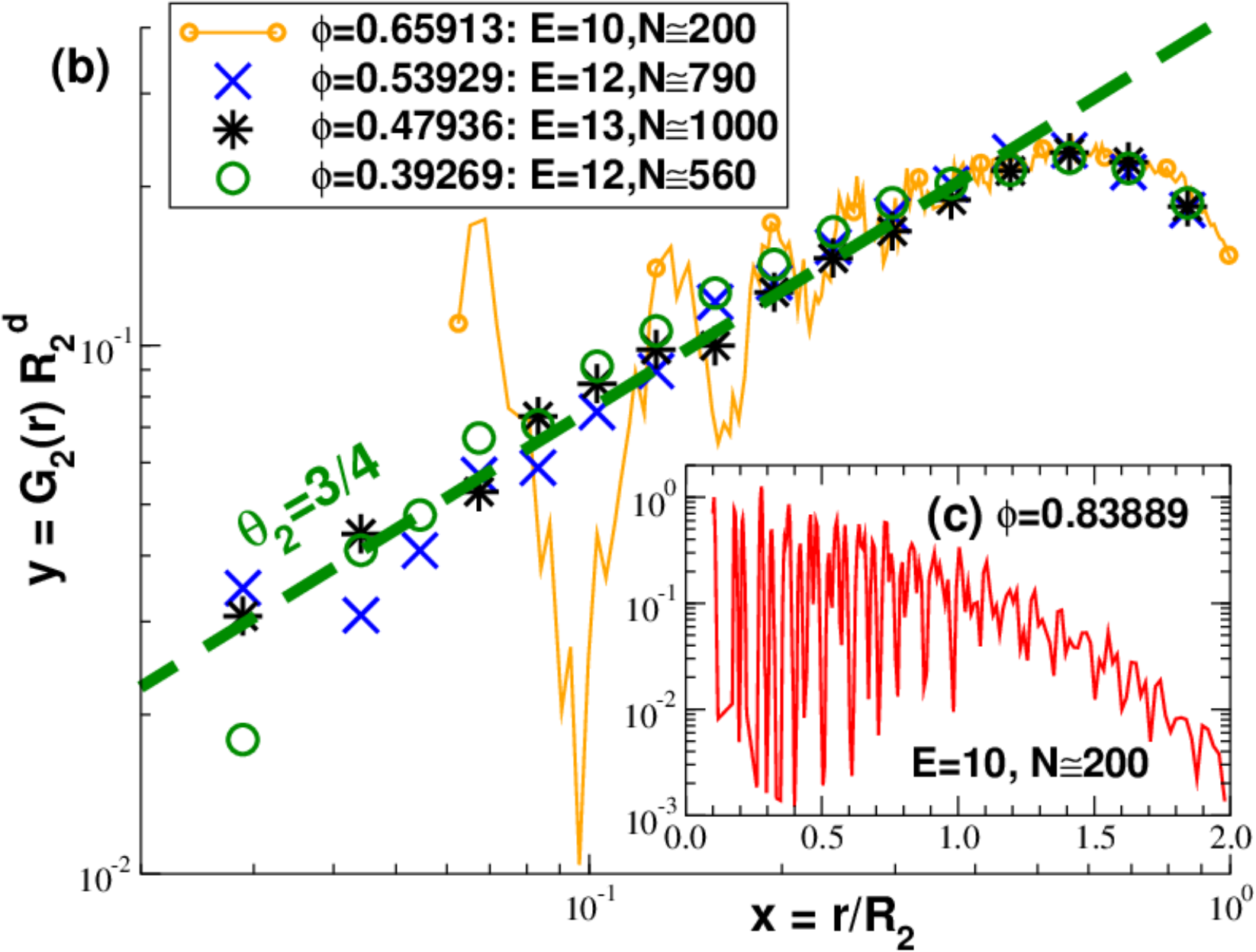}}}
\caption{Rescaled size distributions for contact exponents in the dense limit:
{\bf (a)} data for $i=0$, $i=1$ and $i=2$ focusing on $\phi=0.47936$,
{\bf (b)} several high densities for $i=2$ showing that $\theta_2=3/4$ holds
for not too large $\phi$ while
{\bf (c)} oscillatory behavior sets in for our largest densities. 
}
\label{fig_Gi_dense}
\end{figure}

%
Numerically more challenging distributions are presented in Fig.~\ref{fig_Gi_dilute}
and Fig.~\ref{fig_Gi_dense}. We investigate here three different size distributions
$G_i(r,s)$ with $i=0,1$ and $2$ of internal chain segments of curvilinear length $s$
with $r$ being the spatial distance between the end monomers of the segments.
All size distributions are normalized, i.e. $\int_0^{\infty} G_i(r,s) 2\pi r \ \ddiff r = 1$,
and the typical root-mean-squared sizes $R_i(s)$ of the segments are given by
the second moments 
\begin{equation}
R_i^2(s) = \int_0^{\infty} r^2 G_i(r,s) 2\pi r \ \ddiff r.
\label{eq_Gi_Ri}
\end{equation}
We denote below by $x=r/R_i$ the reduced segment size. In the case that $R_i(s)$
is (essentially) the {\em only} characteristic size characterizing the segment size distribution,
$R_i(s)^d G_i(r,s)$ must for dimensional reasons be functions of $x$ alone. 
Deviations from this scaling are, obviously, expected for small $r \approx \sigma$
and/or $s \approx 1$, where the discrete monomers must matter, and for intermediate
densities where a finite blob size $\xi(\phi)$ sets an additional scale for the segment sizes.
The limit $r \to \sigma^+$ corresponds to the return probability of the segment.
All distributions discussed below increase as power laws $x^{\theta_i}$ 
with $\theta_i > 0$ for $x \ll 1$. 
Their generic form is thus \cite{DegennesBook}
\begin{equation}
R_i^d G_i(r,s) = x^{\theta_i} \ficut(x)
\label{eq_Gi_generic}
\end{equation}
with $\ficut(x)$ being a cut-off function which decays rapidly for $x \gg 1$
but becomes  a finite constant for $x \ll 1$.
For polydisperse systems of distribution $p(N)$ the above segment size distributions
and typical sizes may additionally depend on the total chain length $N$.
One way to avoid this additional argument would be to keep $s$ constant and
to average over all segments of length $s$ in chains with $N \ge s$. 
Used below is a second possibility where $s$ is imposed to be proportional to $N$
and where we average over all chains of length $N'$ similar to a given $N$. 

%
Following standard notation \cite{DegennesBook,DescloizBook,VanderzandeBook}
$i=0$ refers to the case where the segment comprises the total chain, i.e. where $s=N$.
In other words $G_0(r,s=N)$ characterizes the normalized distribution of the distance $r$
between the ends of chains of length $N$. For monodisperse chains it has been shown
that the associated ``contact exponent" $\theta_0$ according to Eq.~(\ref{eq_Gi_generic})
takes the value $\theta_0=11/14$ in the dilute limit and $\theta_0=3/8$ in the dense limit.
As shown by des Cloizeaux \cite{DescloizBook} the exponent $\theta_0$ and the
susceptibility exponent $\gamma$ discussed in the previous two subsections are equivalent
being simply related by $\gamma = 1 + \nu \theta_0$ \cite{DegennesBook}. 
It is thus numerically much simpler to verify the values of $\theta_0$ in both density limits 
for EPs using the $E$-dependence of the mean chain length $\avN$ (cf.~Fig.~\ref{fig_N}{\bf (a)}). 
A {\em direct} test of these values is indicated by circles 
in Fig.~\ref{fig_Gi_dilute} and Fig.~\ref{fig_Gi_dense}{\bf (a)}
for, respectively, dilute and dense systems. 
To obtain a simple scaling behavior we choose either $N \ll g(\phi)$ in Fig.~\ref{fig_Gi_dilute} 
or $N \gg g(\phi)$ in Fig.~\ref{fig_Gi_dense}.
Unfortunately, only tiny fractions of chains have precisely a length $N$ 
(compared to perfectly monodisperse system of imposed length $N$ and same particle number $n$). 
To obtain an acceptable statistics we thus average over chains of length $N'$ 
in an interval of a couple of percent below and above the chain length $N$ indicated in the legends. 
As emphasized by the dashed-dotted lines we then obtain a reasonable confirmation of the expected exponents. 
Not surprisingly, strong deviations are seen especially for small densities
or small $x$ whenever too small distances $r$ are probed.

%
The case $i=1$ corresponds to the size distribution of segments between
either of the two chains ends and an inner monomer at a curvilinear distance $s=N/4$ 
from the respective chain end.
The corresponding rescaled distributions $G_1(r,s)$ are indicated by triangles.
As above for $i=0$ we average over chains of length $N'$ similar to the indicated $N$
to improve the statistics. The bold solid lines indicate the exponents $\theta_1=5/6$
and $\theta_1=1/2$ expected in both density limits. The agreement is
reasonable albeit deviations are strong again for small reduced
distances $x \ll 0.1$ and this especially for small densities. 

%
The most important case $i=2$ corresponds to the size distributions of 
segments of length $s=N/4$ in the middle of chains of length $N$.
To increase the statistics we average over all such segments with end monomers 
being at least a distance $N/4$ from the chain ends.
Moreover, we additionally average over all chains of length $N'$ in a window around 
the indicated chain lengths $N$. The resulting distributions are represented by
squares in Fig.~\ref{fig_Gi_dilute} and Fig.~\ref{fig_Gi_dense}{\bf (a)}.
The expected power law exponents $\theta_2=19/2$ and $\theta_2=3/4$
for both density limits are indicated for comparison by bold dashed lines.
Agreement between data and expected exponents is observed over nearly an order of magnitude.
This is much better than for the other contact exponents $\theta_0$ and $\theta_1$.
The reason for this is simply that the additional averaging over $N/4$ subchains
in the chain middle strongly increases the statistics.

%
An additional problem is finally emphasized in panel {\bf (b)} and panel {\bf (c)} of Fig.~\ref{fig_Gi_dense}
presenting the reduced distribution of inner chain segments $G_2(r,s=N/4)$ for several large densities.
As can be seen, the data for $x \ll 1$ becomes non-monotonic and with strong oscillations
for our largest surface fractions. This can be reduced but not completely suppressed
using larger bins for the histograms. 
The reason for this striking effect is unfortunately not entirely clear
but it must be ultimately a consequence of the hexagonal disk packing in this density limit. 
While only local fluctuations of the monomers around their lattice positions remain possible,
the connectivity network of the bonded interactions is not frozen in the sense that
scission and recombination events at chain ends still take place with a reasonable acceptance rate.
However, we cannot exclude that these events become strongly correlated preventing an efficient sampling 
of the possible bond networks needed for the precise determination of the distributions $G_i(r,s)$.
This issue should be investigated in future work using in addition to the scission-recombination MC steps 
{\em double-bridging MC moves} \cite{FrenkelSmitBook,BWM04} 
where pairs of bonds of neighboring chain segments are exchanged (subject to the no-closed-loop constraint).
Since this allows the bond reorganization independently of the scission energy $E$
this should allow a much better sampling of the possible connectivity networks at high densities
and especially for large $\avN$.

\subsection{Return probabilities}
\label{sub_fi}

\begin{figure}[t]
\centerline{\resizebox{0.9\columnwidth}{!}{\includegraphics*{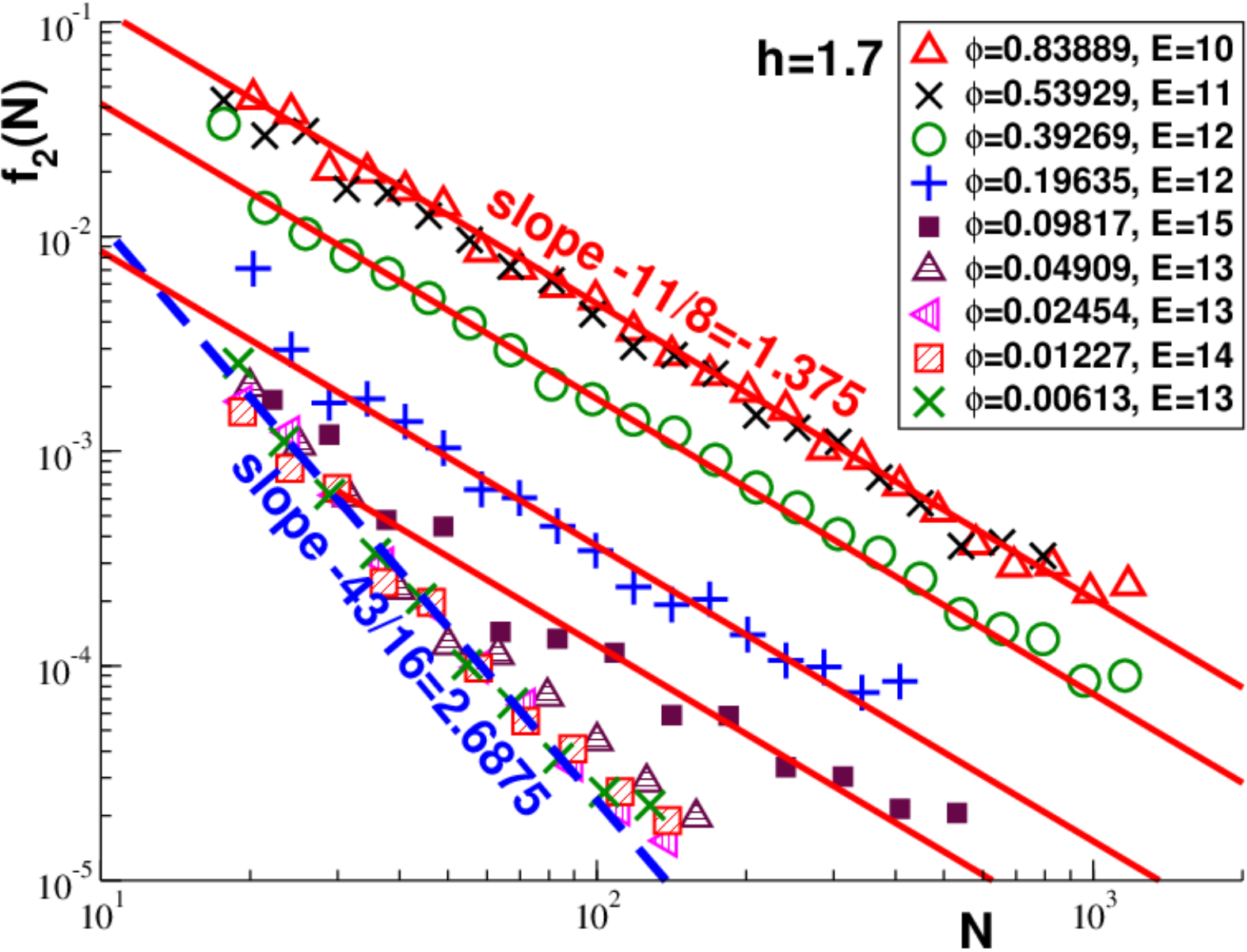}}}
\caption{Fraction of inner monomers in close contact ($r < h=1.7$)
for two monomers being a curvilinear distance $s=N/4$ apart.
Data for a broad range of densities is presented focusing on large $E$.
The dashed line indicates the exponent $-43/16$ for the dilute limit,
the solid lines the corresponding  exponent $-11/8$ for the dense limit.
}
\label{fig_f2_phi}
\end{figure}

Albeit it is impossible to confirm $\theta_2=3/4$ directly from $G_2(r,s)$ for our largest densities,
this exponent can be shown to remain relevant, however, if one takes an appropriate 
average of $G_2(r,s)$ by focusing on the return probability of inner chain segments.
As shown in Fig.~\ref{fig_f2_phi} we compute the fraction $f_2(N)$ of all monomers
in the middle of a chain of total length $N$ between $s=N/4$ and $s=3N/4$ having the other
end of the segment of length $N/4$ at a distance $r < h = 1.7$. 
It follows by integration of Eq.~(\ref{eq_Gi_generic}) that
\begin{equation}
f_2(N) \propto \frac{1}{R_2^d} (h/R_2)^{\theta_2} \propto 1/N^{\nu ( d  + \theta_2)},
\label{eq_Gi_f2N}
\end{equation}
i.e. we expect $f_2(N) \propto 1/N^{43/16}$ in the dilute limit (dashed line) and 
$f_2(N) \propto 1/N^{11/8}$ for dense systems (solid lines).
These exponents are clearly confirmed in Fig.~\ref{fig_f2_phi}
and this even for our largest surface densities. 
Interestingly, for intermediate semidilute densities, say, for $\phi=0.09817$,
a crossover between both power-law limits can be seen.
Similar behavior has been obtained (not shown) for the corresponding fractions
$f_0(N)$ and $f_1(N)$ for the fraction of chain ends being in close
contact with either the opposite chain end or an inner monomer at a curvilinear distance $s=N/4$.

The above characterizations of the contact exponents $\theta_i$ are unfortunately all 
irrelevant for real experimental studies. We turn now to one at least in principle feasible
experimental verification of $\theta_2=3/4$ in the dense limit by means of the form factor $F(q)$ 
\cite{BenoitBook}.

\subsection{Intramolecular form factor $F(q)$}
\label{sub_Fq}
\begin{figure}[t]
\centerline{\resizebox{0.9\columnwidth}{!}{\includegraphics*{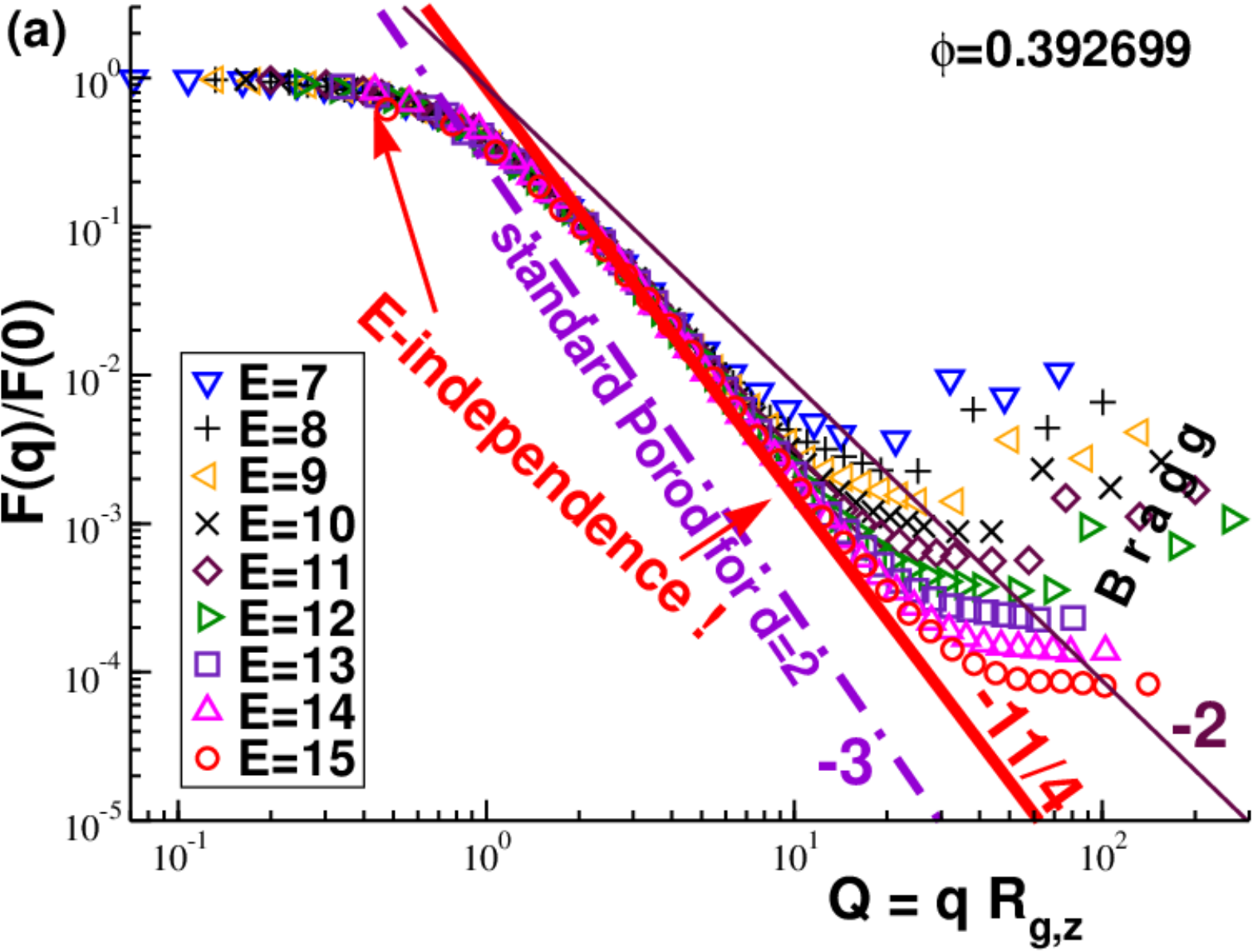}}}
\centerline{\resizebox{0.9\columnwidth}{!}{\includegraphics*{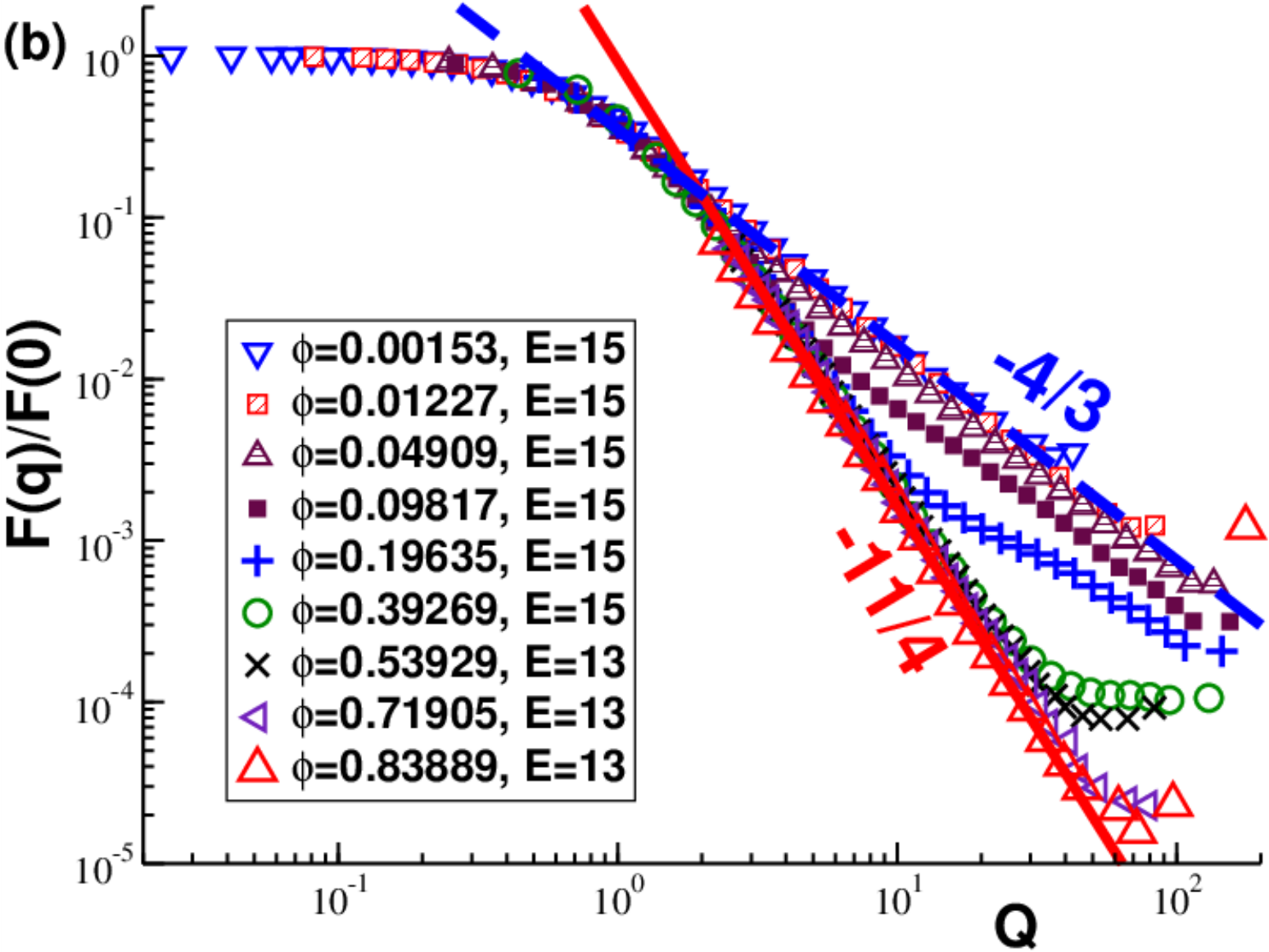}}}
\caption{Rescaled form factor $y=F(q)/F(0)$ as a function
of reduced wavevector $Q= q \Rgz$:
{\bf (a)} data for surface fraction $\phi=0.392699$ and a broad range of $E$
demonstrating that the power-law slope $-11/4$ (bold solid line) expected for large chain lengths and
{\bf (b)} broad range of $\phi$ for one $E$ confirming the expected exponents $-4/3$ (dashed line) and 
$-11/4$ (solid line) for both density limits.
}
\label{fig_Fq}
\end{figure}

%
Conformational properties of polymers or polymer-like aggregates can be determined
experimentally by means of light, small angle X-ray or neutron scattering experiments 
\cite{BenoitBook,DescloizBook}. Using appropriate labeling techniques this allows 
in principle to extract the coherent intramolecular structure (form) factor $F(q)$
\cite{foot_quenched}.
For a given (instantaneous) configuration, wavevector $\qvec$ and chain $c$ of length $N_c$ 
we first compute the sum 
$\sum_{k,l}\exp[i \qvec \cdot (\rvec_k^c-\rvec_l^c)]$
with $\rvec_k^c$ and $\rvec_l^c$ being the positions of two monomers $k$ and $l$ of chain $c$. 
This sum becomes $N_c^2$ for vanishingly small $q \equiv |\qvec|$.
We then sum over all chains $c$ and divide by the total monomer number $n=\sum_c N_c$.
The form factor $F(q)$ presented in Fig.~\ref{fig_Fq} is then obtained
by finally averaging over different wavevectors $\qvec$ of same magnitude $q$
and the $1000$ frames stored. 
%
%
The described normalization is consistent with the limit 
$F(q) \to F(0) = \la N^2 \ra/\la N \ra$ for $q \to 0$
observed in neutron-scattering experiments of polydisperse polymers \cite{BenoitBook}.
In the so-called Guinier regime for small wavevectors
$F(q)$ must scale as \cite{BenoitBook}
\begin{equation}
F(q)/F(0) = 1 - Q^2/d \mbox{ with } Q \equiv q\Rgz
\label{eq_Fq_Guinier}
\end{equation}
being the reduced wavevector and $\Rgz$ the $z$-averaged gyration radius
already considered in Fig.~\ref{fig_RN}.

%
It is well known that the form factor of ``open" (not compact) objects
allows to determine its fractal dimension $\df$ using that
\begin{equation}
F(q) \propto 1/q^{\df} \mbox{ for } 1/R \ll q \ll 1/\sigma
\label{eq_Fq_openobj}
\end{equation}
with $R$ being the typical size of the object and
$\sigma$ a monomer scale (e.g., in our case the disk diameter).
That Eq.~(\ref{eq_Fq_openobj}) holds for our dilute EP systems can be seen for the smaller 
surface fractions $\phi$ indicated in panel {\bf (b)} of Fig.~\ref{fig_Fq}. Tracing $F(q)/F(0)$ 
as a function of the reduced wavevector $Q$ confirms a power-slope with exponent 
$\df = 1/\nuDIL = 4/3$ (dashed line).

Obviously, Eq.~(\ref{eq_Fq_openobj}) does not hold any more if the chains become compact ($\df \to d$), 
i.e., if Porod-like scattering due to the composition fluctuations 
at the surface becomes possible \cite{BenoitBook,foot_perimeter}. 
We have reminded in the Introduction that for compact monodisperse chains
the fractal surface dimension $\dperi$ is related to $\theta_2$, cf.~Eq.~(\ref{eq_dperi}),
and that the form factor decreases for intermediate wavevectors with an exponent $2 \df -\dperi = 11/4$,
cf.~Eq.~(\ref{eq_genPorod}).
These relations must also hold for each chain length $N$ of our polydisperse EPs and 
should remain valid scaling relations if all quantities are replaced by their averages over $N$. 
(We remind that $p(N)$ is given by a simple Schulz-Zimm distribution decaying rapidly
for $N \gg \avN$.) 
We thus expect a data collapse on the asymptotic power law 
\begin{equation}
F(q)/F(0) \propto 1/Q^{11/4}
\mbox{ for } Q \gg 1, q \ll 1/\sigma
\label{eq_Fq_Porod}
\end{equation}
and sufficiently large $\avN$. This is indeed born out by the data given in Fig.~\ref{fig_Fq}.
We present 
in panel {\bf (a)} a broad range of scission energy $E$ for an intermediate surface fraction $\phi=0.392699$
and in panel {\bf (b)} a broad range of $\phi$ for the largest $E$ available.
As shown in panel {\bf (a)} rather large $E$, i.e. large $\avN$, are needed to see
the expected data collapse onto the bold solid lines. This becomes must easier
for larger densities as shown in panel {\bf (b)}. Note that we confirm the
expected exponent for $\phi=0.83889$ over more than an order of magnitude of $Q$.

\subsection{Interchain contact probability $f_3(N)$}
\label{sub_f3}
\begin{figure}[t]
\centerline{\resizebox{0.9\columnwidth}{!}{\includegraphics*{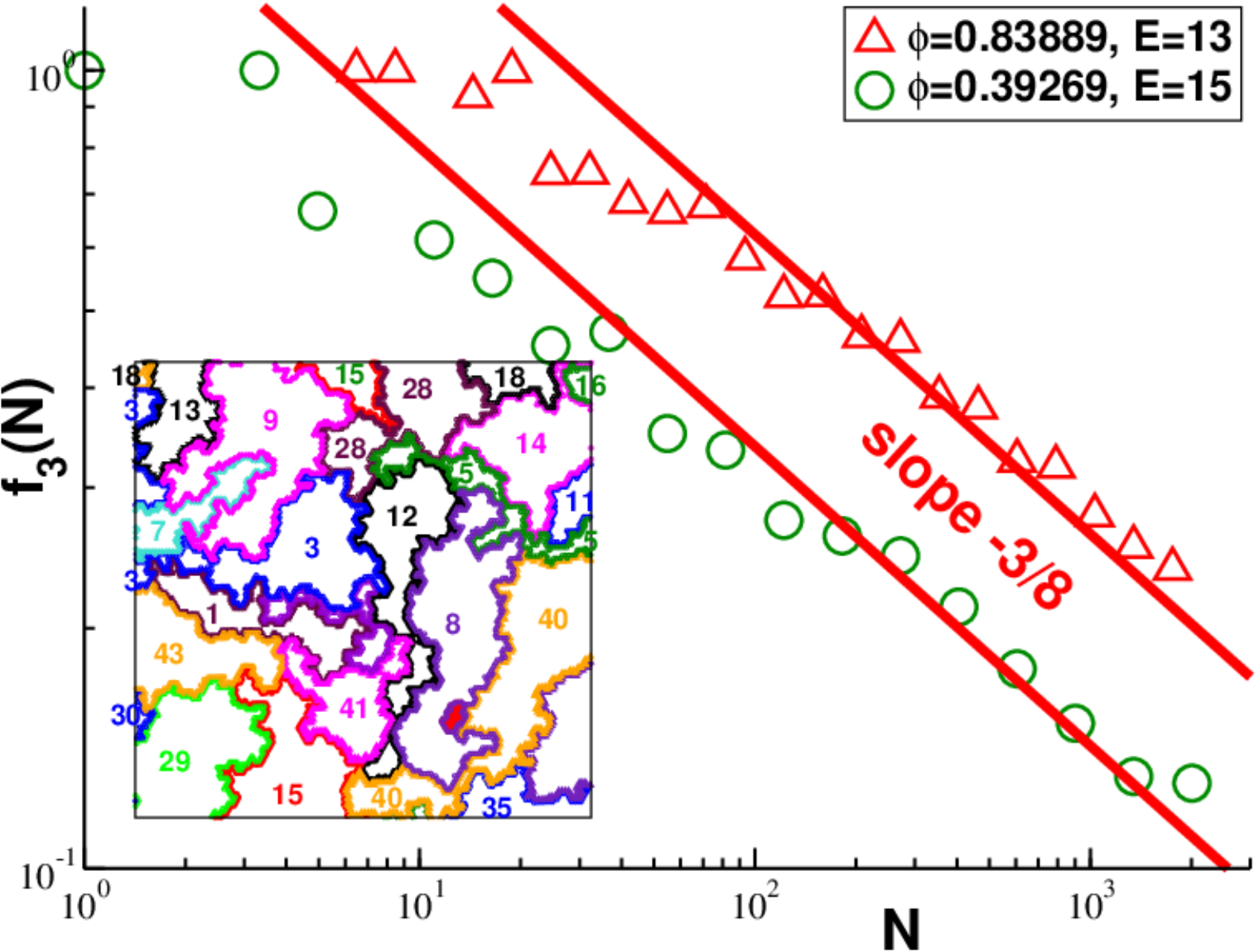}}}
\caption{Fraction $f_3(N)=S(N)/N$ of perimeter monomers,
i.e. monomers with inter chain contacts ($r < h = 1.7$), for two large densities.
The dashed lines indicate the expected exponent $3/8$.
Inset: Snapshot of the perimeter monomers of a configuration at $\phi=0.83889$ and $E=13$.
The numbers indicate chain numbers used for computational purposes.
}
\label{fig_f3}
\end{figure}

Up to know we have entirely focused on {\em intrachain} properties. 
Since the form factor for dense EPs measures the composition fluctuations
at the surface $S(N) \propto R(N)^{\dperi}$ for each chain of length $N$ it is natural
to ask whether it is possible to directly measure the typical perimeter length of our chains. 
We thus compute for all chains of length $N$ the average number of perimeter monomers $S(N)$ 
having at least one monomer from another chain within a distance $h \ll 1.7$.
The perimeter monomers of a subvolume of a configuration at our largest density
are shown in the snapshot given in the inset of Fig.~\ref{fig_f3}.
The perimeter monomer fraction $f_3(N) \equiv S(N)/N$ is expected to scale as 
\begin{equation}
f_3(N) \propto 1/N^{1-\dperi/\df} = 1/N^{\theta_2/\df} = 1/N^{3/8}.
\label{eq_f3_power}
\end{equation}
As shown in the main panel of Fig.~\ref{fig_f3}, 
this is nicely confirmed over two orders of magnitude (bold dashed lines)
for the two indicated $\phi$. We have checked that the precise value of $h$ 
used for identifying a monomer as a surface monomer is irrelevant for the observed exponent
as long as $h \ll R(N)$ for all $N$ probed.

\section{Conclusion}
\label{sec_conc}

\begin{figure}[t]
\centerline{
{\bf (a)}
\resizebox{0.40\columnwidth}{!}{\includegraphics*{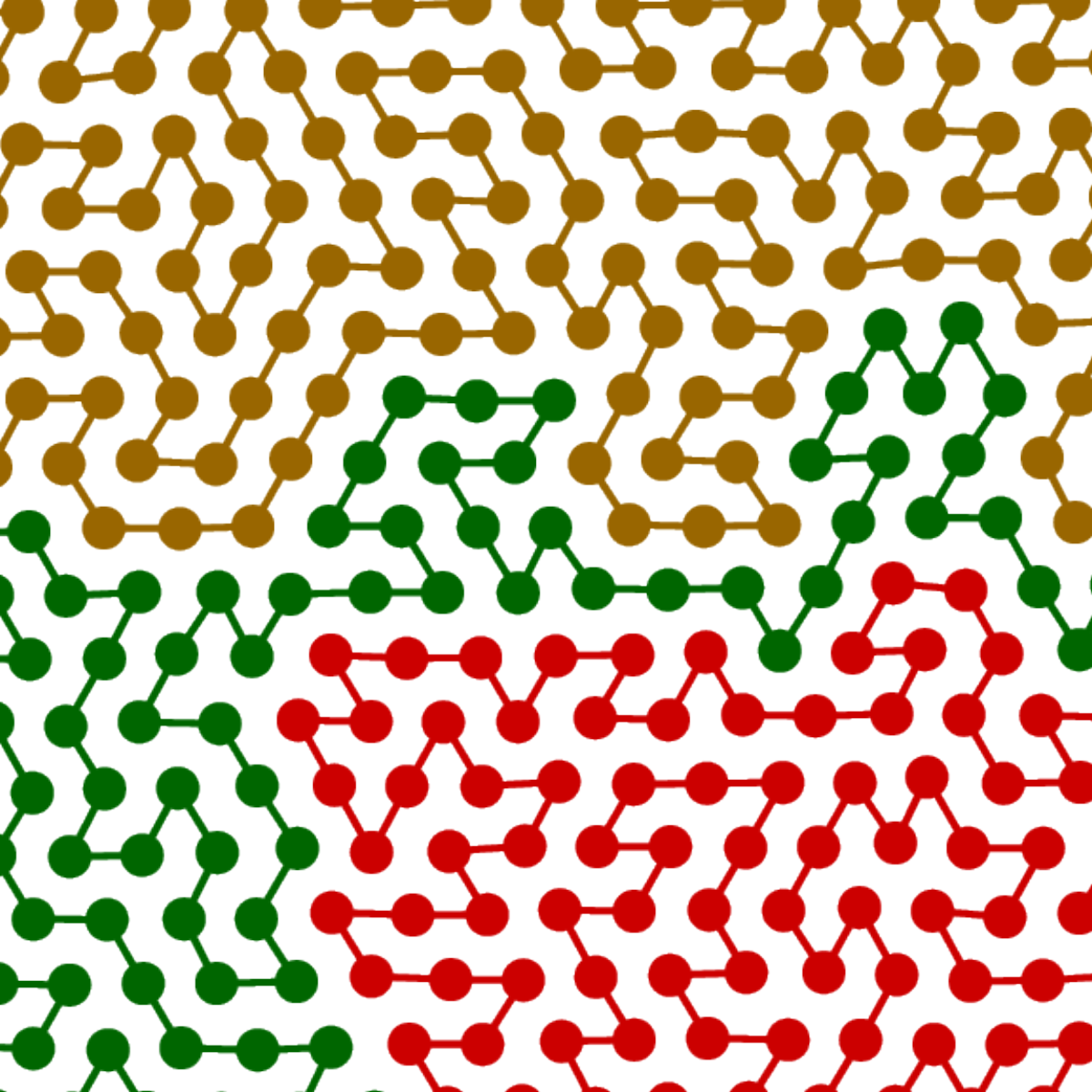}}
\hspace*{0.1cm}
{\bf (b)}
\resizebox{0.40\columnwidth}{!}{\includegraphics*{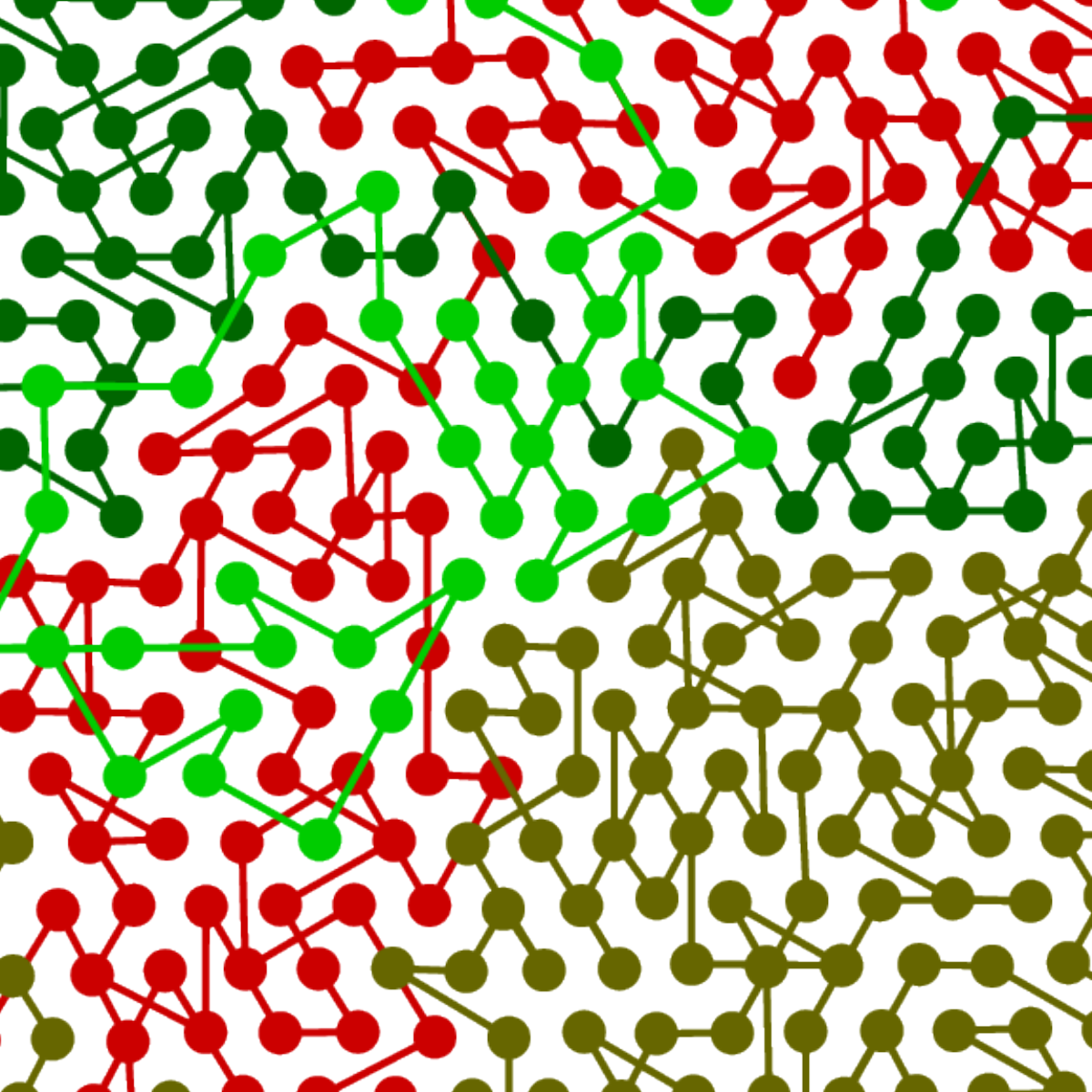}}
}
\vspace*{0.2cm}
\centerline{\resizebox{0.9\columnwidth}{!}{\includegraphics*{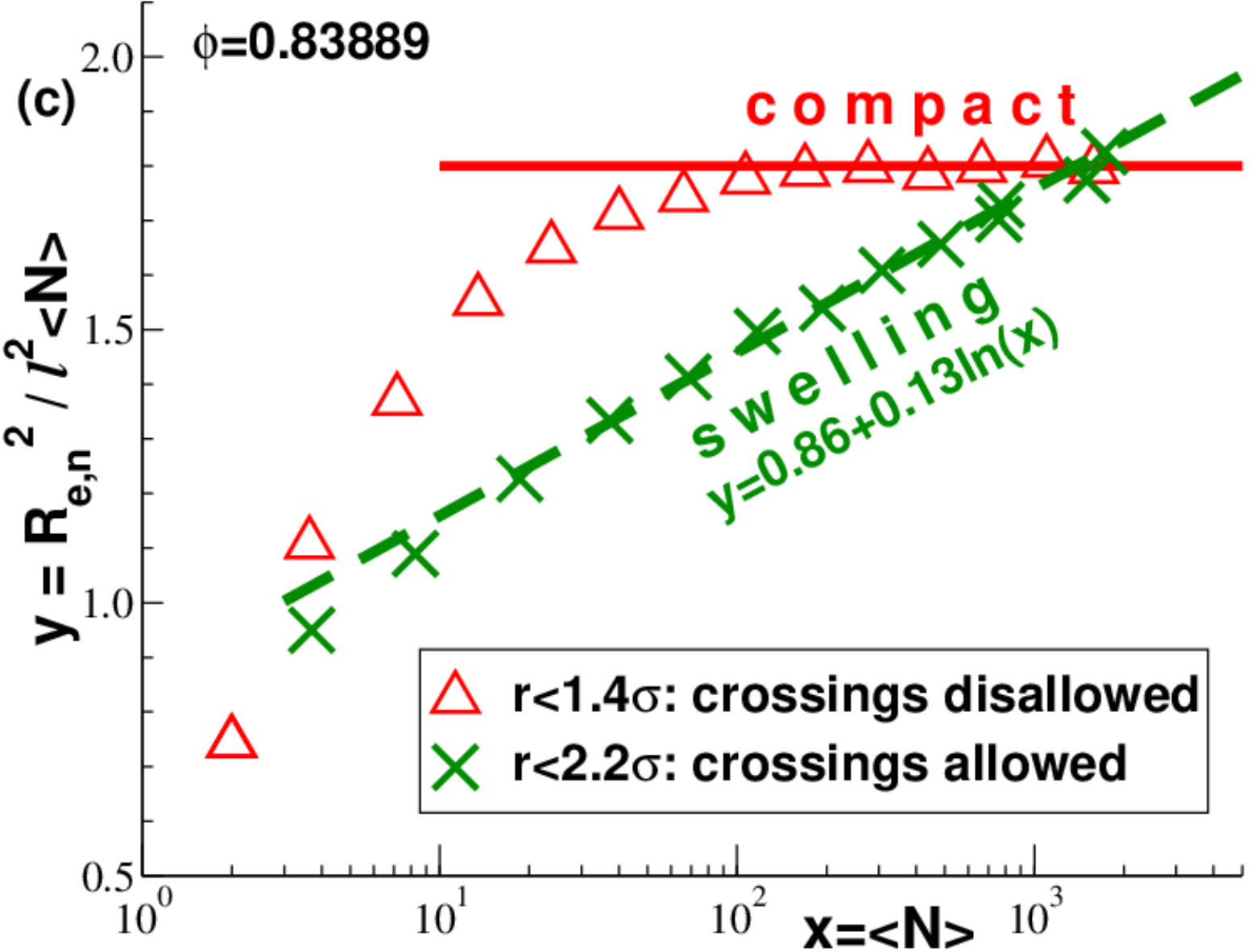}}}
\caption{We have focused in the present study on two-dimensional EPs with disallowed chain crossings 
as shown in panel {\bf (a)} for $\phi=0.83889$.
Sufficiently long chains thus become compact 
as shown by the open triangles in panel {\bf (c)}. However, if instead of
$r < 1.4\sigma$ we impose $r < 2.2 \sigma$ for the allowed distance
$r$ between two bonded monomers chain, crossings become possible
as seen in panel {\bf (b)}.
As shown by the crosses in panel {\bf (c)}, logarithmic swelling is then observed.
}
\label{fig_cross}
\end{figure}

We have investigated in this numerical study conformational static properties of a simple 
generic model system of linear EPs in strictly $d=2$ dimensions. 
Focusing on flexible chains we explicitly disallowed 
the branching of the chains, the formation of closed loops (rings)
and the crossing (intersection) of chain segments (cf. panel {\bf (a)} of Fig.~\ref{fig_cross}).
Theoretically expected universal scaling relations and exponents for asymptotically long chains, 
as summarized in Table~\ref{tab_expo}, have been confirmed both for the dilute limit and, 
more importantly, for semidilute solutions and melts.
Despite of the annealed polydispersity these systems have been seen to be characterized by the same 
asymptotic exponents $\nu$, $\gamma$, $\theta_0$, $\theta_1$ and $\theta_2$
as expected for dense monodisperse chains 
\cite{DegennesBook,Duplantier86a,Duplantier89,CarmesinKremer90,ANS03,MKA09,SMK13}.
For sufficiently large densities $\phi$ and scission energies $E$ 
\begin{itemize}
\item
EPs adopt compact configurations of Flory exponent $\nu=1/2$ (cf.~Fig.~\ref{fig_RN}) and 
\item
the distribution $p(N)$ is set by the susceptibility exponent $\gamma=19/16$ 
(cf. solid lines in Fig.~\ref{fig_cN}),
\item
the mean chain length scales as $\avN \propto \phi^{\alpha} \exp(\delta E)$ with 
$\delta = 16/35$ for systems above the bold dashed line in Fig.~\ref{fig_phiE},
$\alpha = 3/5$ in the semidilute regime ($10 \ll g(\phi) \ll \avN$) and
$\alpha \approx 1$ for larger densities,
\item
the measured intrachain contact exponents $\theta_i$ are reasonably close to the 
predicted values with especially $\theta_2=3/4$ being nicely confirmed
(cf.~Fig.~\ref{fig_Gi_dense} and Fig.~\ref{fig_f2_phi}),
\item
the perimeter of compact EPs have a fractal surface dimension $\dperi=5/4$, cf.~Eq.~(\ref{eq_dperi}),
which determines according to Eq.~(\ref{eq_genPorod}) the generalized Porod decay of the 
intramolecular form factor presented in Fig.~\ref{fig_Fq}.
\end{itemize}
Basically, due to the exponential decay $p(N) \simeq \exp(-\gamma N/\avN)$ in all density regimes
only the behavior of chains of typical length $N \approx \avN$ appears to matter for all 
relevant moments. The properties of EP systems are thus from the scaling point of view
similar to those of monodisperse chains of length $\avN$. The well-known swelling of very long
chains of length $N \gg \avN^2$ in the dense limit \cite{DegennesBook,DescloizBook} thus 
becomes completely irrelevant.
Most importantly, the Flory-Huggins approximation Eq.~(\ref{eq_FH_functional}) appears to remain valid 
for all densities and using Eq.~(\ref{eq_FH_fend_MEL}) holds surprisingly even for extremely
large densities, i.e. possible correlations of the lengths of adjacent chains are irrelevant.

%
As pointed out at the end of Sec.~\ref{sub_Gi}, future numerical work should use 
in addition to the scission-recombination events used in the present study 
double-bridging bond-exchange MC steps \cite{FrenkelSmitBook,BWM04} 
to improve the sampling in the large density limit. 
Moreover, further investigations
should focus on dynamical properties of these systems and on relaxing some of the 
imposed constraints motivated here by theoretical and computational considerations but which 
may not be justified for real experimental systems. 
Let us note first that a finite persistence length of the EPs is only expected to be a crucial parameter
altering the scaling predictions at high surface fractions as soon as it causes some (local or global) 
nematic ordering.
More importantly, we simulated strictly two-dimensional ``self-avoiding walks" without monomer overlap 
and chain intersections as shown in panel {\bf (a)} of Fig.~\ref{fig_cross}. 
This constraint can readily be relaxed by allowing a larger distance $r$ between two bonded monomers, 
say, $r < 2.2 \sigma$ instead of $r < 1.4 \sigma$ as assumed above.
As may be seen from the snapshot in Fig.~\ref{fig_cross}{\bf (b)},
chain crossings then become possible. 
As shown in panel {\bf (c)} for $\Ren^2/l^2\avN$ as a function of $\avN$,
this changes dramatically the asymptotic scaling of the chain sizes:
instead of a compact behavior (horizontal solid line) a {\em logarithmic} swelling (dashed line) is observed.
This agrees with theoretical \cite{ANS03} and numerical \cite{CMWJB05} studies 
on monodisperse chains in ultrathin slits of finite width. Details have yet to be worked out, however,  
concerning the logarithmic deviations for the other power-law relations discussed in the present work.
Another straightforward generalization of the presented algorithm consists
in allowing closed loops to be formed for chains larger than a lower cutoff $N_c$.
For $1 \ll N_c \ll \avN$ rings should thus dominate \cite{Milchev93_Potts}.
A self-similar structure of smaller rings and small linear chains confined within larger rings
is expected which should cause the swelling of rings of all lengths $N \gg N_c$.
Recent results on active-fluid flows in model networks \cite{Bartolo24,Bartolo24b},
which have been mapped on a solid-on-solid model,
suggest that such a self-similar structure should lead to a Flory exponent 
$\nu \approx 3/4$ just as for dilute linear chains.
A direct numerical verification of this claim by means of simulations of EPs is warranted.


\vspace*{0.2cm}
\paragraph*{Acknowledgments.}
We are indebted to O.~Benzerara (Strasbourg) for helpful discussions.

\vspace*{0.2cm}
\paragraph*{Author contribution statement.}
JPW wrote the paper benefitting from contributions of all authors.
 
\vspace*{0.2cm}
\paragraph*{Data availability statement.}
Data sets are available from the corresponding author on reasonable request.

\clearpage
\newpage

\begin{thebibliography}{47}
\bibitem[{foo({\natexlab{b}})}]{foot_blob}
\bibinfo{note}{The blobs are assumed to be much larger than
  the monomer size and much smaller than the overall chain sizes. For $\avN/g
  \to \infty$ the blob size $\xi \propto g^{\deltaDIL}$ thus does not depend on
  the specific polydispersity of a system.}
\bibitem[{foo({\natexlab{c}})}]{foot_alphaD3}
\bibinfo{note}{For semidilute EPs in $d=3$ the corresponding exponents are
  $\delta=1/2$ and $\alpha \approx 0.67$
  \cite{Cates88,CC90,Schaefer92,Schoot97} as reminded in the last column of
  Table~\ref{tab_expo}. These values have been numerically confirmed using a
  lattice polymer model \cite{WMC98a,WMC98b}. For giant micelles there has been
  controversy concerning the value of $\alpha$ due to indirect evidence from
  viscosity measurements being interpreted via the reptation-reaction model
  \cite{CC90} suggesting $\alpha \approx 1.2$
  \cite{Schurtenberger93,Schurtenberger96}. Much larger values $\alpha \approx
  1$ have indeed been observed for a bead-spring off-lattice model
  \cite{MWL00b}, however, at large densities where the semidilute blob scaling
  trivially breaks down.}
\bibitem[{foo({\natexlab{d}})}]{foot_phihigh}
\bibinfo{note}{Our largest surface fraction $\phi=0.83889$ is larger than the
  surface fraction $\phi \approx 0.72$ for the liquid-hexatic transition of
  monodisperse disks (without bonds) but smaller than $\phi \approx 0.898$ for
  a random close packing \cite{Zaccone22} and $\phi = \pi/\sqrt{12} \approx
  0.91$ for the densest hexagonal packing.}
\bibitem[{foo({\natexlab{f}})}]{foot_quenched}
\bibinfo{note}{This assumes that the annealed EPs can be quenched, say, by
  means of a co-solvent preventing further scission and recombination processes
  and that some of these quenched chains can be appropriately labeled.}
\bibitem[{foo({\natexlab{g}})}]{foot_perimeter}
\bibinfo{note}{The surface of a compact object in $d=2$ is more commonly called 
its ``perimeter" \cite{MandelbrotBook,MKA09,MSZ11}.}
\end{thebibliography}

\end{document}